\pdfoutput=1 

\documentclass[a4paper, 11pt, final]{article}

\usepackage{amssymb} 
\usepackage{amsthm} 
\usepackage{amsmath,empheq}
\usepackage{bbm}

\DeclareMathAlphabet{\mathcal}{OMS}{cmsy}{m}{n}
\DeclareMathAlphabet\mathbfcal{OMS}{cmsy}{b}{n}

\DeclareFontFamily{U}{dutchcal}{\skewchar\font=45 }
\DeclareFontShape{U}{dutchcal}{m}{n}{<-> s*[1.0] dutchcal-r}{}
\DeclareFontShape{U}{dutchcal}{b}{n}{<-> s*[1.0] dutchcal-b}{}
\DeclareMathAlphabet{\mathcald}{U}{dutchcal}{m}{n}
\SetMathAlphabet{\mathcald}{bold}{U}{dutchcal}{b}{n}
\DeclareMathAlphabet\mathcalz{T1}{pzc}{mb}{it}

\usepackage[nice]{nicefrac} 
\usepackage{siunitx} 

\usepackage{newtxtext, newtxmath} 

\usepackage{anyfontsize}
\usepackage[utf8]{inputenc}
\usepackage[T1]{fontenc}
\usepackage{microtype} 

\providecommand{\JEL}[1]{\textit{\textbf{JEL: }} #1}
\providecommand{\keywords}[1]{\textbf{\textit{Keywords--- }} #1}

\usepackage{setspace} 

\usepackage{parskip} 
\usepackage{etoolbox}
\usepackage{titlesec}
\titleformat{\section}{\normalfont\Large\bfseries}{\thesection}{1em}{}
\titleformat{\subsection}{\normalfont\large\bfseries}{\thesubsection.}{1em}{}
\titleformat{\subsubsection}{\normalfont\normalsize\itshape}{\thesubsubsection.}{1em}{}

\usepackage{abstract}

\renewenvironment{abstract}
 {\normalfont
  \begin{center}
  \bfseries \abstractname\vspace{-.5em}\vspace{0pt}
  \end{center}
  \list{}{
    \setlength{\leftmargin}{0cm}%
    \setlength{\rightmargin}{\leftmargin}%
  }%
  \item\relax}
 {\endlist}

\usepackage{authblk} 
\usepackage[bottom]{footmisc} 

\usepackage{multicol} 

\usepackage{enumitem} 

\usepackage{graphicx} 
\usepackage{float} 
\usepackage{subcaption}
\usepackage{afterpage} 

\usepackage{printlen}

\usepackage[labelfont=bf]{caption}
\usepackage{color, colortbl}
\definecolor{LightGray}{rgb}{0.93,0.914,0.914}    
\usepackage{soul} 
\usepackage{longtable,rotating} 
\usepackage{booktabs} 
\usepackage{multirow} 
\usepackage{arydshln} 
\usepackage{bigdelim} 

\usepackage[most]{tcolorbox}            

\PassOptionsToPackage{hyphens}{url} 

\usepackage{fancyhdr}
\fancyheadoffset{0cm}
\makeatletter
\newcommand{\quickwordcount}[1]{
  \immediate\write18{texcount -quiet -incbib -sub=none -utf8 -1 -sum -merge -encoding=utf8 #1.tex > #1-words}%
  \immediate\openin\somefile=#1-words
  \read\somefile to \@@localdummy
  \immediate\closein\somefile
  \setcounter{wordcounter}{\@@localdummy}
  \@@localdummy
}
\makeatother

\usepackage{silence}
\usepackage[style=apa,backend=biber,natbib,hyperref]{biblatex}
\DeclareLanguageMapping{british}{british-apa}
\defbibenvironment{bibliography}{\enumerate}{\endenumerate}{\item}

\usepackage[colorlinks=true,allcolors=gray]{hyperref} 
\let\orgautoref\autoref

\renewcommand{\autoref}[1]
{%
\def\equationautorefname{Eq.}%
\def\sectionautorefname{Sec.}%
\def\subsectionautorefname{Subsec.}%
\def\figureautorefname{Fig.}%
\def\subfigureautorefname{Fig.}%
\orgautoref{#1}%
}

\usepackage[nameinlink,capitalise]{cleveref}

\usepackage{algorithm,algpseudocode}

\makeatletter
\newlength{\trianglerightwidth}
\algnewcommand{\LineCommentCont}[1]{\Statex \hskip\ALG@thistlm%
  \parbox[t]{\dimexpr\linewidth-\ALG@thistlm}
{\leftskip=\algorithmicindent
  \hangindent=\algorithmicindent 
  \hangafter=1%
  \strut\makebox[\algorithmicindent][c]{$\triangleright$}#1\strut}
  } 
\makeatother

\usepackage[left=1.8cm, right=1.8cm, bottom=2.5cm, top=2.5cm]{geometry}

\begin{document}


\renewcommand{\figureautorefname}{Fig.}
\onehalfspacing



\newcommand{\MainTitleText}{An extendable, integrated, and dynamic approach to forecasting and stress-testing credit risk}

\title{\fontsize{20pt}{0pt}\selectfont\textbf{\MainTitleText
}}


\author[,a]{\large Marcel Muller \thanks{ ORC iD: 0000-0002-8585-2426; email: \url{marcelcelliers@gmail.com}}}
\author[,a,b]{\large Arno Botha \thanks{ ORC iD: 0000-0002-1708-0153; email: \url{arno.spasie.botha@gmail.com}}}
\author[,a,b]{\large Conrad Beyers \thanks{ ORC iD: 0000-0002-2870-4718 ; email: \url{conrad.beyers@up.ac.za}}}
\affil[a]{\footnotesize \textit{Department of Actuarial Science, University of Pretoria, Private Bag X20, Hatfield, 0028, South Africa}}
\affil[b]{\footnotesize \textit{National Institute for Theoretical and Computational Sciences (NITheCS), Stellenbosch 7600, South Africa}}
\renewcommand\Authands{, and }

    

\makeatletter
\renewcommand{\@maketitle}{
    \newpage
     \null
     \vskip 1em%
     \begin{center}%
      {\LARGE \@title \par
      	\@author \par}
     \end{center}%
     \par
 } 
 \makeatother
 
 \maketitle

{
    \setlength{\parindent}{0cm}
    \rule{1\columnwidth}{0.4pt}
    \begin{abstract}
    An integrated and extendable approach for stress-testing loan portfolios is presented, which includes both a loan production component and a credit risk component. In this approach, we simulate a completed portfolio using realistic loan parameters and distributional assumptions. Thereafter, we generate the uncertain cash flow history of these loans within a multistate probabilistic framework. We illustrate our approach using a simulation-based study, though the approach can be fit to real-world data. Such a simulation-based approach is ideal for stress-testing since it allows for evaluating a range of conditions. From these completed loans, we compute portfolio-level credit risk metrics, e.g., default and loss rates. Stress scenarios are introduced by varying the loan parameters accordingly within a broader Monte Carlo setup, thereby resulting in a range of portfolios. A classical approach to stress-testing does not typically integrate loan production or embed the correlation structure amongst risk metrics. In our approach, we integrate the forecasting of risk metrics with receipt-generation. Given data, the loan parameters within our extendable approach can be dynamically modelled as functions of input variables using any applicable technique. Overall, our approach can render predictions that are more dynamic and flexibly tuned, which can enhance stress-testing practices within any bank.
    \end{abstract}
     
    \keywords{Stress-testing; Credit Risk; Loan Production; Macroeconomic Stress; Simulation.}
     
     \JEL{C44, C63, G21.}
    
    \rule{1\columnwidth}{0.4pt}
}

\subsection*{Acknowledgements}
\noindent We sincerely thank Prof. Jonathan Crook for his review of our work and subsequent contributions, which have greatly enhanced this manuscript. 
This work is not financially supported by an institution or study grant, and has no conflicts of interest that may have influenced the outcome of this work.

\noindent Word count (excluding front matter):  7511 



\newpage

\section{Introduction}
\label{sec:ch1}

To mitigate its credit risk, banks hold loss reserves to absorb potential future credit losses. These loss reserves can be partitioned into three distinct parts, where the first is the minimum amount called \textit{loss provisions} (or simply \textit{provisions}) that should be held for offsetting \textit{expected losses} (ELs). The requirements for provisions are specified by the \textit{International Financial Reporting Standards} (IFRS) 9, as published by the \citet{ifrs9_2014}. A second loss reserve called \textit{regulatory capital} is held over and above the provisions and is used to absorb \textit{Unexpected Losses} (ULs) transpiring from extreme and/or unforeseen events, such as domestic economic recessions. ULs are governed by a set of standards developed by the \citet{basel2019}, known as the Basel framework, which sets regulatory capital requirements. See \S 2.5--2.56 of \citet{botha2021Proc} for an in-depth explanation of ELs and ULs. Finally, if the bank were to breach its regulatory capital requirements during even more adverse scenarios, such as a global recession or a pandemic, an additional loss reserve called \textit{stressed capital} must be held, as required by \S 33.8 of \citet{basel2019}. To estimate this capital requirement, the bank must define a severe but plausible scenario to shock its future credit losses and then compare the resulting capital estimates with the minimum regulatory requirements. This forecasting procedure is known as a \textit{stress-test} and forms part of the bank's \textit{internal capital adequacy assessment process}, as described in \S 30.2 of \citet{basel2019}. Such a stress test is ultimately used to assess the going concern (or solvency) of the bank in the event of adverse/stress scenarios.

The procedure used in forecasting the loss reserves for a particular credit risk portfolio, and its associated balance sheet, comprises a few components. \citet{matthew2004}, \citet{foglia2008}, and \citet{ferrari2011} described a general multistage framework for such a forecasting task using a stress-testing lens. Firstly, a coherent and specific systematic scenario needs to be defined that affects the repayment behaviour of all obligors and therefore drives the overall loss within a specific portfolio. The factors that constitute this scenario are known as the \textit{stressors}. Secondly, the stressors need to be linked to the risk parameters of a risk model in sensitising its estimates to the systematic stress. These risk parameters include the \textit{probability of default} (PD), \textit{loss given default} (LGD), and the \textit{exposure at default} (EAD). The product of these parameters constitute "credit risk", i.e., the potential credit loss associated with lending out funds, as discussed by \citet[\S 1]{botha2021Proc}. The stressed risk parameters are then used to evaluate the impact of the scenario on the loss provisions through some type of risk/loss measure, e.g., the ELs, ULs, default rates, etc.

The correlation structure amongst the risk parameters of the credit risk model should preferably be preserved during the stress-test such that the stress scenario manifests realistically and intuitively on future loss reserves. Such correlation structures have been shown to exist in literature, with \citet{frye2000CollateralDA}, \citet{frye2000CollateralD},  and \citet{Jokivuolle2003IncorporatingCV} showing the prevalence and effects of omitting these parameter correlations. We call the incorporation of the correlation structure between the risk parameters an \textit{integrated} approach to stress-testing, where its application is likely to yield a more accurate estimation of the true underlying credit risk within the portfolio subject to the stress scenario. Regarding existing literature, \citet{sorge2006}, \citet{breuer2012}, \citet{bellotti2013}, and \citet{bocchio2023} stressed the PD, although all assumed the LGD and EAD to be constant. \citet{rosch2007} formulated an extended framework to stress both the PD and its correlation to macroeconomic factors (known as the asset correlation) simultaneously, but did not go as far as to stress the EAD and LGD. \citet{jokivoulle2013} stressed both the PD and LGD whilst accounting for the correlation between these two risk parameters through a Cholesky decomposition. To the best of our knowledge, the only literature that has considered all three risk parameters simultaneously in credit stress testing is that of \citet{jimenez2009}. The authors modelled the number of defaults in a portfolio using a time-series model with macroeconomic variables as inputs, whilst loan-level LGDs are obtained by drawing deviates from a time-invariant Beta distribution. Moreover, EADs of performing loans are forecast by drawing deviates from a Gamma distribution, whilst the EADs of non-performing loans are forecast with deviates from an inverse Gaussian distribution. Both EAD distributions are constructed by modelling their parameters using macroeconomic variables as inputs. Lastly, the correlation structure amongst the credit risk parameters is preserved indirectly through the correlations of the individual risk parameters to the macroeconomic variables. This approach cannot be regarded as fully `integrated' since the LGD parameter is not stressed and the interdependence among risk parameters is not directly taken into account.

Alternatively, receipt (or cash flow) forecasting frameworks, which forecast future account states, can be used to infer all credit risk parameters such that they are inherently integrated through a single quantity; i.e., the humble cash flow at some point in time. These forecasting frameworks are often formulated on the basis of profit-margin assessments and do not generally consider all required account states and transitions for estimating credit losses. For example, \citet{djeundje2025} developed a multi-state framework for stressing the profit margins of loans by forecasting their future cashflows/receipts. The receipts are naturally dependent on account states, where the transition probabilities between these states are modelled as a function of various inputs, including macroeconomic variables like the unemployment rate and inflation. However, the framework considers the default state as absorbing, thereby potentially omitting recurring default events and eventual write-offs such that any estimated loss rate will be inaccurate. That said, our study has some overlap with this work regarding the formulation of our stress-testing approach in that we also use a multistate framework.

Most stress-tests lack an explicit loan production component that generates new advances (or loans) over the forecasting horizon, which unrealistically assumes a static balance sheet. The inclusion of such a loan production component, or a self-styled \textit{extendable} component, would ensure greater realism and accuracy within the ensuing stress-tests. \citet{jimenez2009} acknowledges this shortcoming and explained that the credit market will likely grow during economic upturns and contract during downturns. This implies the need to stress the loan production component itself, thereby expanding the stress-testing framework. The authors subsequently incorporated a loan production component in their framework that forecasts the number of new loans that are periodically disbursed using a time-series model with macroeconomic variables as covariates. They concluded that this component is significant in forecasting the loss distribution and recommend its incorporation when forecasting credit risk. One may certainly extend this component to simulate not only the number of new loans, but also the attributes of each, such as its principal amount and annual interest rate. These attributes reflect the loan's inherent risk and may certainly be powerful predictors when modelling its associated PD, LGD, and EAD. To the best of our knowledge, no existing literature incorporates such a comprehensive loan production component, which suggests a gap that our work can fill.

Credit risk forecasting is generally performed at the portfolio-level in the interest of simplicity, which may impose restrictions on the availability, quantity, and quality of data. However, portfolio-level modelling limits the scope of realistic forecasts since it cannot account for most of the dynamic and idiosyncratic phenomena that affect a typical loan, as discussed by \citet{crook2010}. In contrast, \citet{smith1995forecasting}, \citet{ferrari2011}, \citet{grimshaw2011markov}, and \citet{botha2026multistate} have all
demonstrated that loan-level modelling can easily incorporate such phenomena towards obtaining a more realistic set of loss predictions. We refer to the use of loan-level models as a \textit{dynamic} approach to stress-testing.
Using a residential mortgage portfolio, \citet{Gaffney2014} formulated a "transitions-based PD-model" that can produce loan-level cash flow forecasts as a function of state-dependent input variables, e.g., policy and unemployment rates, and buy-to-let status. Their self-styled "loan loss forecasting" model estimates various types of annual loan-level cash flows, ultimately obtaining expected losses. This forecasting model relies in turn on two \textit{intensity models} that predict the instantaneous risk of either defaulting or curing when given inputs, as originally built by \citet{kelly2016good} using Irish data.
Another notable example is the work of \citet{bocchio2023}, who forecast the underlying loan-level behaviour as a function of factors that drive the credit risk model for a retail mortgage portfolio. Aspects that are forecast over the forecasting horizon include the borrower-specific balance, interest, property value, and loan-to-value ratio. Our study too shall incorporate a transition-based approach to forecasting elements of credit risk.

Given these gaps, we introduce a simulation-based stress-testing approach in which a scenario-dependent series of risk-weighted receipts (or cash flows) are produced for each loan over its forecast lifetime. In so doing, we simultaneously and indirectly forecast all of the credit risk parameters, and then stress them by tweaking the systematic scenario under which receipts are generated. Although this stress-testing approach and its constituent components can certainly be fit to a particular dataset, we apply it within a simulation framework to allow us to investigate various real-world scenarios without depending on a particular dataset. A simulation-based approach also offers a certain degree of versatility in testing a range of conditions, including extreme ones.
We further assess the uncertainty and variability of each scenario by casting our stress-testing approach within a Monte Carlo setup such that a range of portfolios can be generated per scenario. By focussing on receipt-generation, our approach is said to be 'integrated` since it embeds the correlation structure amongst the risk parameters implicitly via the receipts. Furthermore, a loan production component is formulated within this broader forecasting framework, thereby integrating the advancing of new loans into the stress-test and rendering the approach as `extendable'. Lastly, our approach is considered as `dynamic' since it enables some of its parameters to be modelled as time-dependent idiosyncratic factors at the loan-level.

In summary, our approach is best described as an \textit{extendable, integrated, and dynamic forecasting and stress-testing} (EIDFAST) approach and is formulated theoretically in \autoref{sec:ch2}. The EIDFAST-approach is then calibrated to a baseline and a stressed scenario in \autoref{sec:ch3} towards generating a range of realistic loan portfolios for each scenario. Having forecast the various credit risk metrics indirectly for each of these portfolios, we discuss the results in \autoref{sec:ch4} and finally conclude the study in \autoref{sec:ch5}. Moreover, we illustrate how the EIDFAST-approach is parametrised for a specific scenario within \autoref{app:app01}, followed by an illustration and comparison of the evolutions of the EIDFAST-approach's parameters under two scenarios in \autoref{app:app02}. The risk metrics used in assessing the effects of the two scenarios on the simulated loan portfolios are lastly detailed in \autoref{app:app03}. The R-based codebase used for this study is provided and maintained by \citet{muller2026KasmeerSourcecode}.

\section{A transition-based method for generating receipts: the EIDFAST-approach}
\label{sec:ch2}

\begin{figure}[ht!]
    \centering
    \includegraphics[width=0.8\linewidth,height=0.32\textheight]{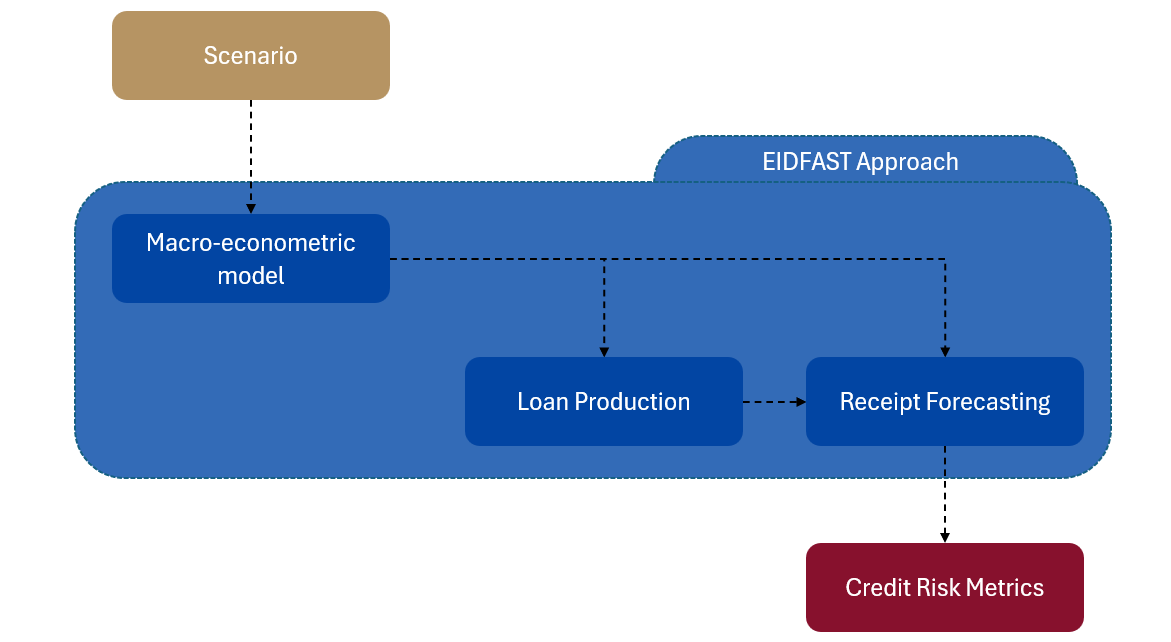}
	\caption{An enhanced framework for forecasting the credit risk metrics of a particular credit portfolio.}
	\label{fig:General_Framework}
\end{figure}

We propose a new method consisting of two major components: a loan production component and a receipt forecasting component. This novel EIDFAST-approach directly models the repayment behaviour of obligors by forecasting their associated receipts/cash flows over time. In so doing, the portfolio-level credit risk metrics, i.e., PDs and LGDs, can be indirectly calculated from these forecast receipts (which may include zeros).  In contrast, a classical stress-testing approach would ordinarily forecast these various metrics directly, though it would be unable to predict repayment behaviour at the account-level and at scale. The novel approach is illustrated in \autoref{fig:General_Framework}.

For the loan production component, each  loan $i$ is simulated over the period $t \in \{1,\dots,n_i,n_i+1,\dots,n_i+m\}$, where
$\{1,\dots,n_i\}$ represents the historical period and $\{n_i+1,\dots,n_i+m\}$ constitutes the forecast horizon (or stress period). This simulation is achieved using a two-step process. In the first step, the number of new monthly advances (or loan volumes) is simulated for each period by drawing random deviates from a particular distribution.
The second step involves simulating three characteristics of each new loan. This includes the principal amount $B_0$, where the outstanding balance at $t$ is denoted as $B_t$.
The remaining two characteristics are the original term $T$, and the nominal monthly interest rate $r$. 
Each of these three characteristics are independently drawn from different distributions, as will be discussed later in \autoref{sec:ch3}. The instalment (or expected payment) at $t$ is calculated as
\begin{equation} \label{eq:Instalments}
    I_t = B_0 \cdot \frac{r(1+r)^T}{(1+r)^T-1} \, .
\end{equation}

Within the credit risk component, receipts are forecast in two steps over $t$.
In the first step, a random loan account assumes an \textit{account state}, denoted as $Y_t$ at each $t$.
The account state $Y_t$ is then used in the second step to infer the associated receipt $R_t\geq0$ (or cash flow receivable) of the account over $t$. 
This $Y_t$ is itself generated over $t$, whereafter $R_t$ is duly determined at each $t$; all of which will be discussed later in this section.
This forecasting exercise is repeated for all sampled obligors in the portfolio and across all loan periods to produce a series of account state forecasts and an associated series of receipts. The series of forecast receipts subsequently enables the calculation of the associated balances and arrears of each account at each $t$. These balances and arrears can then be used to measure the delinquency level such that period-specific PDs and LGDs can be calculated to obtain forecasts of the portfolio-level credit risk, and therefore the loss provision.

For simplicity, the state space of each $Y_t$ can assume one of the following states: payment (P), delinquency (D), settlement (S), and write-off (W), i.e., $Y_t \in \{\mathrm{P,D,S,W}\}$. An account will be in the payment state (or up-to-date) when it is not behind on its instalments and has not yet been settled or written-off. Conversely, an account will be in the delinquency state when it is behind on its instalments but has not yet been settled or written-off. Settlement occurs when an account has repaid its last expected balance, irrespective of timing. Lastly, an account will be written-off after accruing a certain level of arrears or, equivalently, having spent an significant amount of time in D. States P and D are transient states in that accounts may transition between these two states before ultimately progressing to one of the two absorbing states S or W, from which they cannot escape.

In following \citet[\S 1]{norris1997markov}, the finite sequence of forecast payment states $Y_{n + 1}, \dots, Y_{n + m}$ is assumed to be a homogenous first-order Markov chain $(Y_t)_{t\geq 0}$ over discrete-time $t\in\mathbb{Z}_{\geq0}$. 
We assume stationarity since non-homogeneity is enforced through varying the parameters of the EIDFAST-approach over the forecast horizon, as discussed in \autoref{sec:ch3}.
As a random process, $(Y_t)_{t\geq 0}$ may be estimated from data, especially since each loan history denotes a \textit{sample path} from the underlying Markov chain. The $4\times4$ transition matrix $M$ that governs $(Y_t)_{t\geq 0}$ has entries $p_{kl}$ that denote the transition probability from state $k$ to $l$, i.e., $p_{kl}=\mathbb{P}\left( Y_{t} = l \, | \, Y_{t-1}=k \right)$ between any discrete (monthly) time points $t-1$ and $t$, as illustrated in \autoref{tbl:TransitionMatrix}. From \citet{anderson1957}, the maximum likelihood estimates (MLEs) of each $p_{kl}$ is $n_{kl} / n_k$, where $n_{kl}$ is the total number of observed transitions from $k$ to $l$, and $n_k$ is the number of transitions starting in $k$. Additionally, it is reasonably assumed that accounts start in the payment state, i.e., $Y_1=\mathrm{P}$.

\begin{table}[ht!]
\caption{A conceptual transition matrix showing possible transitions amongst four payment states that a typical loan can assume during its lifetime that is measured discretely (which is typically monthly).}
\centering
\label{tbl:TransitionMatrix}
\begin{tabular}{c l c c c c}
\multicolumn{2}{c}{} & \multicolumn{4}{c}{To} \\
\multicolumn{2}{c}{} & P & D & S & W \\
\cline{3-6}
\multirow{4}{*}{\rotatebox[origin=c]{90}{From}} 
  & P & \multicolumn{1}{|c}{$p_{\mathrm{PP}}$} & $p_{\mathrm{PD}}$ & $p_{\mathrm{PS}}$ & \multicolumn{1}{c|}{$p_{\mathrm{PW}}$} \\
  & D & \multicolumn{1}{|c}{$p_{\mathrm{DP}}$} & $p_{\mathrm{DD}}$ & $p_{\mathrm{DS}}$ & \multicolumn{1}{c|}{$p_{\mathrm{DW}}$} \\
  & S & \multicolumn{1}{|c}{0}                 & 0                 & 1                 & \multicolumn{1}{c|}{0} \\
  & W & \multicolumn{1}{|c}{0}                 & 0                 & 0                 & \multicolumn{1}{c|}{1} \\
\cline{3-6}
\end{tabular}
\end{table}

The account state $Y_t$ is used in the second step to condition the forecast of the receipts. Let the receipt be $R_t \geq 0$, which denotes the net cash flow that is received from a borrower at $t$ whose account is in state $Y_t$. In forecasting $R_t$, we are interested in using the \textit{payment probability} (PP) of an account at time $t$, i.e., $\mathbb{P}(R_t\geq I_t)$, as estimated by $b_t$; itself a placeholder for an eventual risk model. Note the distinction between $b_t$ and the transition probability $p_{kl}$.
An account in state P at $t$ will have $b_t=1$ and it is assumed (for simplicity) that the associated receipt is exactly $R_t=I_t$. Similarly, if $Y_t=\mathrm{S}$, then settlement suggests $b_t=1$. Since $B_{t-1}$ relies on previous receipt forecasts, $B_t$ is stochastic. We therefore forecast the receipt for settlements as
\begin{equation} \label{eq:Receipt_Settlement}
    R_t = B_{t-1} \cdot (1+r) \, .
\end{equation}
In this case, the receipt is simply equal to $B_{t-1}$, accrued with one month's interest using the associated loan-specific annual interest rate $r$.

Given the transition P$\rightarrow$D of an account between times $t-1$ and $t$, no payment would have been received and hence the PP and receipt are both certain, i.e., $b_t=0$ and $R_t=0$. If this account spends any additional periods within delinquency, it can either delve deeper into delinquency by continuing to miss payments, or it can maintain the same delinquency level by paying only the expected instalment amount at that time. The receipts are thus forecast as
\begin{equation} \label{eq:Receipts_Perf}
    R_t = \begin{cases} 
      I_t \quad & \text{if} \ u < b_t \\
     0 \quad & \text{otherwise}
    \end{cases} \, \quad \text{for } Y_t=\mathrm{D},
\end{equation}
where $u$ is a random deviate drawn from a uniform distribution.

In rendering a sensible forecast of a write-off event, we must realistically ensure that an account destined for write-off spends an appropriate amount of time within the implicit default state.
Let $g_0(t)$ denote the number of payments in arrears (which is a typical delinquency measure) at some point $t$, as investigated by \citet{botha2021}. Define a default threshold $d\geq 0$ such that default occurs when $g_0(t)\geq d$ for some $t$ and $d=3$ payments in arrears, i.e., an implicit default "state". This time of default will be referred to as $t_d$. Then, let $\tau$ be a random integer-valued \textit{sojourn time} that represents the time spent in default before progressing to the write-off state W, hence the time of write-off is given as $t_w = t_d + \tau$.
Furthermore, an account is deliberately kept in D until $\tau$ periods have lapsed, i.e., $Y_t=\mathrm{D}$ for $t\in[t_d,t_w-1]$. This constraint aligns with practice where write-off occurs only some time after the default time $t_d$.
Conditional on $t < t_w$ and $Y_t=D$, we define another PP-estimate $b_t'<b_t$, where $b_t'=\mathbb{P}(R_t \ge I_t | Y_{t}=D)$. The associated receipts during $t\in[t_d,t_w-1]$ are now forecast as  
\begin{equation} \label{eq:Receipts_Del}
    R_t = \begin{cases}
      I_t \quad & \text{if} \ u < b_t' \\
     0 \quad & \text{otherwise}
    \end{cases} \, .
\end{equation}
The PP $b_t'$ in \autoref{eq:Receipts_Del} is deliberately different to the $b_t$ in \autoref{eq:Receipts_Perf} in reflecting the worsening credit risk as the account moves closer to write-off.
Upon reaching $t_w$, the associated receipt is assumed to be a relatively large non-zero amount that reflects the collection efforts made during the workout process of the defaulted loan. However, an amount is expected to be lost, and so let $w$ denote the proportion of the outstanding balance $B_{t_w}$ that is to be written-off. The associated receipt is then forecast as 
\begin{equation} \label{eq:Receipts_WOff}
    R_{t_{w}} = B_{t_{w}-1} \cdot \left(1 + r \right) \cdot \left(1 - w \right) \, .
\end{equation}
Lastly, the receipts of accounts that have entered either of the absorbing states (S or W) are zeroed for the remainder of the forecast horizon, i.e., $R_{t>t_{s}}=0$ for $Y_{t_s}=\mathrm{S}$ at the time of settlement $t_s$, and $R_{t>t_{w}}=0$ for $Y_{t_w}=\mathrm{W}$.
\section{Calibrating the method for stress-testing a generated loan portfolio}
\label{sec:ch3}


All data in this study, i.e., new loan volumes, loan attributes, and loan receipts, are carefully simulated based on logic and industry insights. The initial parametrisation of our method reflects standard/normal operating conditions, which serves as a \textit{baseline} scenario against which to compare, and is detailed in \autoref{sec:ch3.1}. After simulating a portfolio under this baseline scenario across an artificial historical period, the portfolio is subjected to a \textit{stress scenario} in which the parameters of the method are recalibrated so that they reflect the effects of an adverse systemic event (e.g., a macroeconomic downturn). The parametrisation of the stress scenario is detailed accordingly in \autoref{sec:ch3.2}.

\subsection{Parametrising our framework towards generating a baseline portfolio}
\label{sec:ch3.1}



New loans are generally extended by banks at all times, subject to various attributes that depend on macroeconomic conditions. In our case, we assume a \textit{baseline} scenario that reflects stable systematic conditions and serves as a base for a stress-testing exercise. The loan production component is subsequently constructed as follows. New loan volumes are simulated for each period by drawing random deviates from a truncated Normal distribution with a mean of 500, a standard deviation of 10, and a lower truncation point of 5. These parameters are set simply to promote computational efficiency when generating the broader loan portfolio.
We assume that the principal amount $B_{0,i}$ of each newly disbursed loan $i=1,\dots,N$ follows a truncated right-skewed Beta distribution. The $\alpha$-shape parameter is set to 2 and the $\beta$-shape parameter is set to 5, whilst the lower and upper truncation points are respectively set to the currency amounts of R 20,000 and R 2,000,000. We believe these parameters sufficiently embed the principle of risk-based pricing regarding loan amounts of a mid-market mortgage portfolio.
For loan terms, a truncated normal distribution is selected, and random deviates are drawn to simulate each loan's contractual term $T_i$. The mean of the distribution is sensibly set to 240 months, representing the typical term in a mortgage portfolio, with the standard deviation set to 10, and upper and lower truncation points respectively set to 120 and 360.
Lastly, annual interest rates $r_i$ are simulated by drawing random deviates from a truncated right-skewed Beta distribution such that the majority of accounts have lower rates, which again reflects risk-based pricing. The $\alpha$-shape parameter is set to 2 and the $\beta$-shape parameter is set to 5. Zero interest rates are avoided by bounding the minimum of the distribution to 5\%, whilst unrealistically high rates are prevented by bounding the maximum to 20\%. These bounds are guided by Regulation 42(1) of the National Credit Act Regulations, as published by the South African government in the \citet{govgazette2015interest}, which stipulates that the maximum interest rate charged to a South African mortgage is no more than 12\% above the current policy rate (with the policy rate typically varying between 3.5\% and 8\%).
The distribution of each of these loan characteristics is shown within \autoref{app:app01}.


In forecasting receipts, the transition matrix from \autoref{tbl:TransitionMatrix} is set to \autoref{tbl:TransitionMatrix_Est}, whose values are rationalised as follows. The majority of loans spend their lifetimes in the payment state P, with only a few ever transitioning into D, hence the high $\mathrm{P} \rightarrow \mathrm{P}$ and low $\mathrm{P} \rightarrow \mathrm{D}$ probabilities. Those loans transitioning into $\mathrm{D}$ tend to stay there for short periods, with the largest portion thereof returning to the performing state, as reflected by the low $\mathrm{D} \rightarrow \mathrm{D}$ and high $\mathrm{D} \rightarrow \mathrm{P}$ probabilities. Since contractual loan terms span several years, the probability of a loan settling is relatively low at all times, and even lower for an account in delinquency since the outstanding balance is larger due to amounts in arrears. Moreover, the proportion of accounts that are written-off is also low by design, lest the generated loan losses exceed any profit. This rhetoric is embedded throughout the small absorbing transition probabilities $P \rightarrow S$, $P \rightarrow W$, $D \rightarrow S$ and $D \rightarrow W$.
Furthermore, the risk development of delinquent accounts is assumed to be moderately high over all time, with accounts only repaying one month's instalment 40\% of the time, i.e., $b_t = 40\%$ for all $t$. Delinquent accounts destined for write-off are bound to have deteriorating payment probabilities such that a write-off definition is triggered. We assume that this repayment performance is half that of the `normal' delinquent accounts, i.e., $b_t' = 20\%$. These settings reflect industry experience in that the eventual risk metrics (e.g., default rates) behave sensibly.

\begin{table}[ht!]
\caption{The account state transition matrix used for simulating transitions during the historical period consisting of discrete time points $t_1,\dots,t_n$, where the values are experimentally set using expert judgement.}
\centering
\label{tbl:TransitionMatrix_Est}
\begin{tabular}{c l c c c c}
\multicolumn{2}{c}{} & \multicolumn{4}{c}{To} \\
\multicolumn{2}{c}{} & P & D & S & W \\
\cline{3-6}
\multirow{4}{*}{\rotatebox[origin=c]{90}{From}} 
  & P & \multicolumn{1}{|c}{0.987} & 0.010 & 0.002 & \multicolumn{1}{c|}{0.001} \\
  & D & \multicolumn{1}{|c}{0.351} & 0.643 & 0.0015 & \multicolumn{1}{c|}{0.0045} \\
  & S & \multicolumn{1}{|c}{0.0}   & 0     & 1     & \multicolumn{1}{c|}{0} \\
  & W & \multicolumn{1}{|c}{0}     & 0     & 0     & \multicolumn{1}{c|}{1} \\
\cline{3-6}
\end{tabular}
\end{table}

Also contributing to the eventual loss of a loan is the length of time $\tau_i$ that a written-off loan $i$ spends in default. The longer this time, the higher the probability that the bank will receive at least some receipts. From experience, a write-off decision typically occurs anytime between 1 and 48 months in default during the workout period, depending on the jurisdiction and legal processes. This insight is embedded in our simulation by simulating $\tau_i$ through drawing random deviates from a truncated and right-skewed negative Binomial distribution; thereby ensuring integer-valued sojourn times. 
The probability parameter of this distribution is set to 0.2, the size/shape parameter is set to 3, and the lower and upper bound is respectively set to 0 and 48. These settings again reflect industry experience in that the distributional shape appears realistic.
Lastly, the most substantial factor contributing to the eventual loss is the portion $w$ of the outstanding loan balance that is written-off at the write-off point $t_w$. The majority of write-offs tend to be small amounts since the collateral, being the financed property, can be auctioned off to recover most of the outstanding balance. Random deviates from a truncated right-skewed Beta distribution are therefore drawn to emulate these smaller loss percentages. The $\alpha$-shape parameter is set to 1.5, the $\beta$-shape parameter is set to 3, and the lower and upper bounds are respectively set to 0.05 and 0.8 to ensure a sensible write-off percentage, as aligned with industry experience.
Each of the sojourn time and write-off rate distributions, as described here for the baseline scenario, are illustrated in \autoref{app:app01}.
\subsection{Incorporating a specific stress scenario in forecasting the loan portfolio}
\label{sec:ch3.2}


In accordance with industry experience, stress scenarios typically elicit a strong response from a bank that results in stricter lending criteria.
In our stress scenario, all simulation components related to loan production are stressed by methodically varying the parameters of each distribution over the forecast period.
Let $\kappa(t)$ be a time-dependent parameter of a specific component within our framework, where $t_1,\dots,t_n,t_{n+1},\dots,t_{n+m}$ are the points in time of the portfolio's lifetime and $n$ is the end of the historical period. For example, $\kappa(t)$ could be the mean used in simulating new loan volumes at each time $t$, or it could be the lower/upper bound of a truncated distribution at $t$.
Though many options exist, we choose a simple linear function for $\kappa(t)$ such that the stress propagates gradually over the stress test horizon, i.e., 
\begin{equation} \label{eq:parameter_stress}
    \kappa(t_i) = \kappa(t_n) + (t_i - t_n)\cdot\Delta \, , \ \text{where} \ \Delta = \frac{\kappa({t_{n+m}) - \kappa(t_{n}})}{m} \ \text{and} \ i \geq n .
\end{equation}

In illustrating our stress-testing approach, we choose the change in its parameters as follows.
New loan volumes are assumed to decrease by 25\% over the stress test horizon, and we therefore reduce the mean parameter of this component by 25\% over the same period.
Similarly, the principal amounts of new loans are reduced by 25\% over the stress test horizon, which we enforce by reducing the upper truncation parameter of this component's Beta distribution by 25\% over the same period.
The annual interest rates charged to new accounts are increased by 25\% over the stress test horizon, thereby reflecting a worsening credit risk environment and associated risk-based pricing. 
This 25\%-value is merely chosen for illustrative purposes, though future work can certainly refine its selection.
This stress is embedded by increasing both the lower and upper truncation bounds of the component's distribution by the same amount over the same period.
No stress is induced on contractual terms of new loans since they are likely to be scenario-invariant.
All other parameters are kept unchanged from the historical period over the forecast horizon. These stresses on the loan production components are summarised in \autoref{tbl:parameters_stress_summary_loan_prod}, where $\kappa$ represents the parameter of the associated sub-component. The resulting component distributions are compared with those used in the baseline scenario and are discussed in \autoref{sec:ch4}.

\begin{table}[ht!]
\caption{A summary of the stressed parameter values within the loan production components, shown for the start of the stress test horizon $t_n$, and end of the horizon $t_{n+m}$. The marginal changes $\Delta$ are calculated using \autoref{eq:parameter_stress}.}
\centering
\label{tbl:parameters_stress_summary_loan_prod}
\begin{tabular}{l l c c c}
\hline
\textbf{Component}           & \textbf{Parameter ($\kappa$)}              & \textbf{Value at $t_n$} &\textbf{ Value at $t_{n+m}$}    & \textbf{$\Delta$}\\
\hline
Volume              & Mean                   & 500            & 375           & -3.47    \\
Principal amount    & Upper truncation bound & R 2 000 000    & R 1 500 000   & -R 13 888.89    \\
Nominal interest rate& Upper truncation bound & 20\%           & 25\%          & 0.14\%    \\
Nominal interest rate& Lower truncation bound & 5\%            & 6.25\%       &  0.04\%    \\
\hline
\end{tabular}
\end{table}

Components relating to the receipt forecasts of the simulated loans are stressed similarly to the loan production components.
For account state migrations, each transition rate within the transition matrix, as shown in \autoref{tbl:TransitionMatrix_Est}, is gradually changed using \autoref{eq:parameter_stress} from its original value at $t_n$, until it reaches its pre-defined value at the end of the stress horizon $t_{n+m}$. This operation renders the transition matrix as a time-dependent construct. These transition rates from state $k$ to $l$ at time $t_{n+m}$ are set to the values in \autoref{tbl:TransitionMatrix_Est_Stress}, again according to expert judgement. Compared to the historic period, performing accounts during the forecast horizon are logically more likely to transition into $D$ and $W$ given the induced stress. For the same reason, delinquent accounts are more likely to remain in $D$ and transition into $W$.

\begin{table}
\caption{The account state transition matrix defined at the end of the stress-test horizon $t_{n+m}$, contrasted with the starting matrix from \autoref{tbl:TransitionMatrix_Est}. Interleaving matrices between the start and end matrices are obtained using \autoref{eq:parameter_stress}.}
\centering
\label{tbl:TransitionMatrix_Est_Stress}
\begin{tabular}{c l c c c c}
\multicolumn{2}{c}{} & \multicolumn{4}{c}{To} \\
\multicolumn{2}{c}{} & P & D & S & W \\
\cline{3-6}
\multirow{4}{*}{\rotatebox[origin=c]{90}{From}} 
  & P & \multicolumn{1}{|c}{0.97466} & 0.02094 & 0.0014  & \multicolumn{1}{c|}{0.003} \\
  & D & \multicolumn{1}{|c}{0.33431} & 0.65114 & 0.00105 & \multicolumn{1}{c|}{0.0135} \\
  & S & \multicolumn{1}{|c}{0.0}     & 0       & 1       & \multicolumn{1}{c|}{0} \\
  & W & \multicolumn{1}{|c}{0}       & 0       & 0       & \multicolumn{1}{c|}{1} \\
\cline{3-6}
\end{tabular}
\end{table}

Receipts of delinquent accounts are stressed as follows. For delinquent accounts not destined for write-off, we gradually lower $b_t$ using \autoref{eq:parameter_stress} from 40\% at $t_n$ to 35\% at the end of the stress period.
Conversely, receipts of accounts that are destined for write-off are stressed via three components, again using \autoref{eq:parameter_stress}. Firstly, the value of $b_t'$ for these accounts is gradually lowered using \autoref{eq:parameter_stress} from 20\% to 15\%. Since fewer payments are likely to be paid, write-off amounts are likely to be higher. Secondly, the sojourn time distribution is stressed so that accounts are likely to spend more time in delinquency, thereby increasing the likelihood of paying more receipts (although with a lower likelihood) before being written-off. Specifically, the upper truncation bound of the sojourn time distribution is increased by 12 months, whereas the negative Binomial distribution's skew is reduced by scaling the size parameter upwards by 25\%. Lastly, the write-off rate is increased since recoveries are likely to deteriorate during the stress scenario. The associated write-off rate distribution is adjusted to be left-skewed by scaling the $\beta$-shape parameter downwards by 60\%.
The net effect of these three stressed components is expected to result in increased losses of written-off loans, as we shall demonstrate in \autoref{sec:ch4}. The application of these stresses onto these parameters are summarised in \autoref{tbl:parameters_stress_summary_receipts}.

\begin{table}[ht!]
\caption{A summary of the stressed parameter values within the receipt forecast component, shown for the start of the stress test horizon $t_n$ and the end of the horizon $t_{n+m}$, where changes $\Delta$ are calculated using \autoref{eq:parameter_stress}. }
\centering
\label{tbl:parameters_stress_summary_receipts}
\begin{tabular}{l l c c c}
\hline
\textbf{Component} & \textbf{Parameter ($\kappa$)} & \textbf{Value at $t_n$} & \textbf{Value at $t_{n+m}$} & \textbf{$\Delta$} \\
\hline
Payment probability & $b_t$                      & 40\% & 35\% & -0.14\% \\
Payment probability & $b_t'$                      & 20\% & 15\% & -0.14\% \\
Sojourn time               & Size                    & 3    & 3.75 & 0.02 \\
Sojourn time               & Upper truncation bound  & 48   & 60   & 0.33 \\
Write-off rate             & Shape-parameter $\beta$ & 3    & 1.2  & -0.05 \\
\hline
\end{tabular}
\end{table}

\section{Results \& Discussion}
\label{sec:ch4}


In illustrating our approach, we conduct a stress-test by simulating a loan portfolio using the EIDFAST-approach for both our baseline and stress scenarios. For both scenarios, the simulation proceeds by simulating a portfolio over an artificial historical period subject to the set of parameters from the baseline scenario; see \autoref{sec:ch3.1}. The simulation continues with this same set of parameters over the succeeding forecast period for the baseline scenario, but reverts to the set of stressed parameters for the stress scenario; see \autoref{sec:ch3.2} for the specification of the latter.
Both simulations are repeated 50 times within a broader Monte Carlo setup, as determined experimentally. Doing so can help with investigating the degree of uncertainty within each scenario, and with exploring related statistical properties of risk metrics. The loan portfolios resulting from these simulations shall now be compared, whereafter the associated risk metrics will be assessed.

As detailed in \autoref{sec:ch3.2}, the lending criteria is restricted over the forecast horizon for the stress scenario by stressing three loan production sub-components. The resulting distributions are illustrated and assessed in \autoref{app:app02}, which we summarise as follows.
Firstly, the distribution of new loan volumes shifts to the left under the stress scenario such that the number of new monthly loans decreases.
Secondly, the distribution of the principal amounts disbursed to these new loans shifts to the left, again reflecting stricter lending criteria during the stress scenario.
Lastly, the distribution of the assigned nominal interest rates shifts to the right during the stress scenario, thereby charging these new loans with higher nominal interest rates.
The simulation of contractual loan terms is kept unchanged from the baseline scenario since it is unlikely to be affected by any stress scenario.
Having simulated the loan portfolio 50 times, a distribution of each of these stressed sub-components is constructed to analyse the evolution of these sub-components over the forecast horizon. We summarise these linear evolutions in \autoref{tbl:loan_prod_stress_evolution} at various times during the forecast horizon. In line with the distributional shifts, the average number of new loans across all simulations contracts over the horizon. Moreover, the average principal amount and nominal interest rate respectively decreases and increases across all simulations. These results are as expected for our stress scenario.

\begin{table}[ht!]
\caption{Means of the stressed sub-components within the loan production component at multiple periods during the forecast horizon, as subjected to our stress scenario.}
\begin{tabular}{ccccc}
\label{tbl:loan_prod_stress_evolution}
\textbf{Time in forecast horizon} & \textbf{New loans} & \textbf{Principal amounts} & \textbf{Nominal interest rates} & \textbf{Contractual terms} \\ \hline
Last known             & 499                   & R 593 755                           & 10.19 \%                               & 240                           \\
1 month                & 497                   & R 590 158                           & 10.26 \%                               & 240                           \\
3 months               & 492                   & R 582 049                           & 10.39 \%                               & 240                           \\
6 months               & 481                   & R 570 127                           & 10.60 \%                               & 240                           \\
9 months               & 467                   & R 559 654                           & 10.81 \%                               & 240                           \\
12 months              & 458                   & R 547 780                           & 11.01 \%                               & 240                           \\
24 months              & 418                   & R 497 827                           & 11.78 \%                               & 240                           \\
36 months              & 374                   & R 448 954                           & 12.58 \%                               & 240                           \\ \hline
\end{tabular}
\end{table}

The receipt forecasting sub-components are stressed via the change in the payment state transition matrix, the payment probabilities, and the sojourn times and write-off rates that are simulated for accounts that are destined for write-off.
The last two sub-components are simulated by drawing random deviates from specific distributions (see \autoref{sec:ch3.1}), and full analyses of these sub-components are provided in \autoref{app:app02} across the baseline and stress scenarios. We highlight the following salient points of these analyses.
Both the sojourn time and write-off rate distributions shift to the right during the stress scenario, which reflects worsening collection conditions.
To assess the evolution of the sojourn time and write-off rate over the forecast horizon, we subsequently calculate the average value of each simulated sub-component, for each period during the stress horizon across all simulation runs, with respect to each scenario. The evolutions of the sojourn time and write-off rate distributions are linear and summarised in \autoref{tbl:credit_risk_stress_evolution} at various times, where the increase in each sub-component over the horizon reflects the aforementioned distributional shifts.

\begin{table}[ht!]
\caption{The mean sojourn times and write-off rates of loans that have defaulted at multiple periods during the forecast horizon, as subjected to our stress scenario.}
\centering
\begin{tabular}{ccc}
\label{tbl:credit_risk_stress_evolution}
\textbf{Time in forecast horizon} & \textbf{Sojourn times} & \textbf{Write-off rates} \\ \hline
Last known             & 12.24                   & 35.82 \%   \\
1 month                & 12.26                   & 36.15 \%   \\
3 months               & 12.41                   & 37.26 \%   \\
6 months               & 13.10                   & 38.90 \%   \\
9 months               & 13.60                   & 40.69 \%   \\
12 months              & 13.88                   & 42.19 \%   \\
24 months              & 15.96                   & 48.39 \%   \\
36 months              & 17.84                   & 49.60 \%   \\ \hline
\end{tabular}
\end{table}

\begin{figure}[H]
\centering
\includegraphics[width=1\linewidth,height=0.325\textheight]{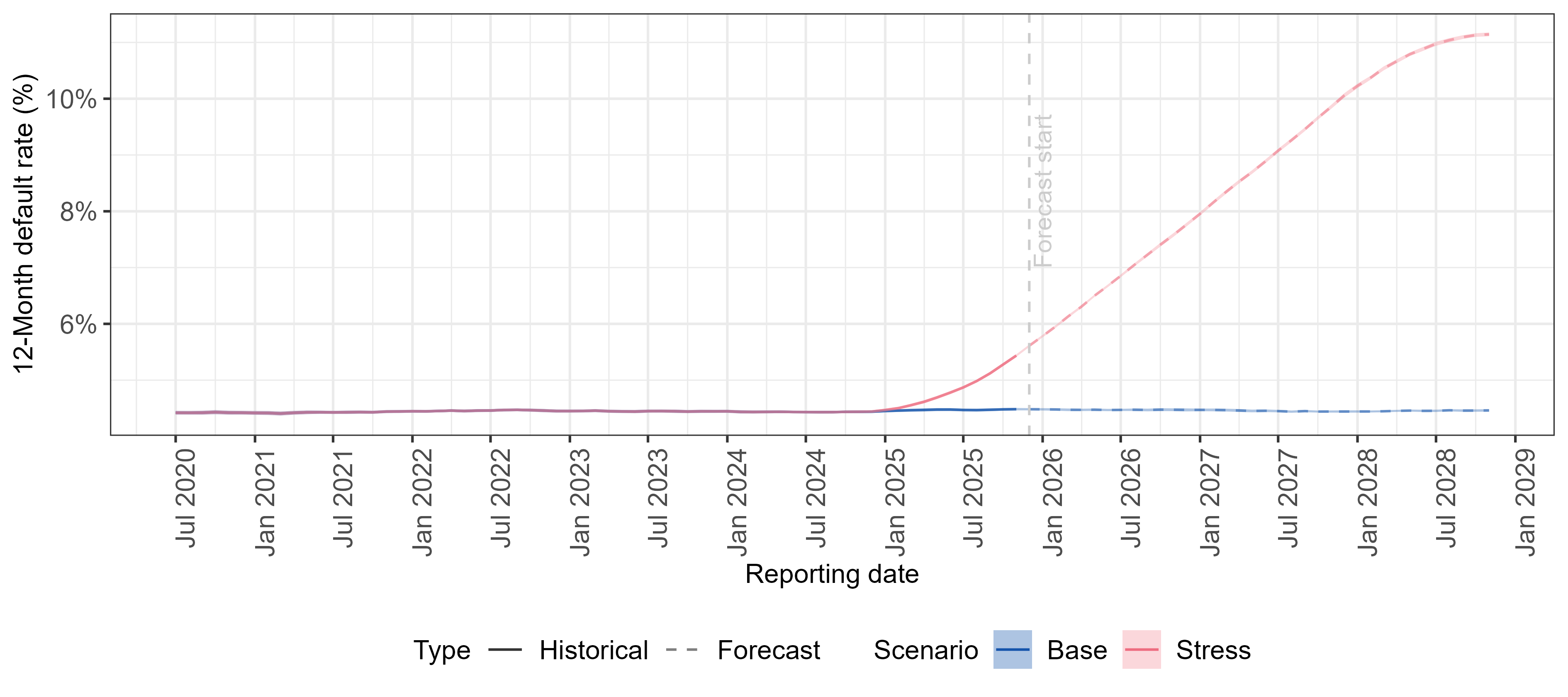}
\caption{A comparison between the average 12-month default rate calculated across all simulated loan portfolios, with respect to the baseline and stress scenario. The bands indicate the 95\% confidence interval.}
\label{fig:12Month_PD_Comparison}
\end{figure}

Having simulated the loan portfolio over historical and forecast periods, we calculate the portfolio-level 12-month default rate and loss rates as our two main risk metrics. Doing so can help with estimating the potential increase in the overall portfolio losses. Each risk metric is mathematically defined within \autoref{app:app03}.
Consider \autoref{fig:12Month_PD_Comparison}, which illustrates the average 12-month default rate, as calculated across all simulation runs for both the historical and forecast periods, with respect to the baseline and stress scenarios. The default rate is stable for the baseline scenario and varies between 3\% and 3.5\%, reflecting the consistency of the framework parameters. In contrast, the default rate of the stress scenario diverges from the start of the forecast period and eventually settles above 11\%. This ~3.5 fold increase reflects a substantial deterioration in the repayment performance of the underlying loans and is reasonably expected for a secured portfolio during adverse events. 

\begin{figure}[H]
\centering
\includegraphics[width=1\linewidth,height=0.325\textheight]{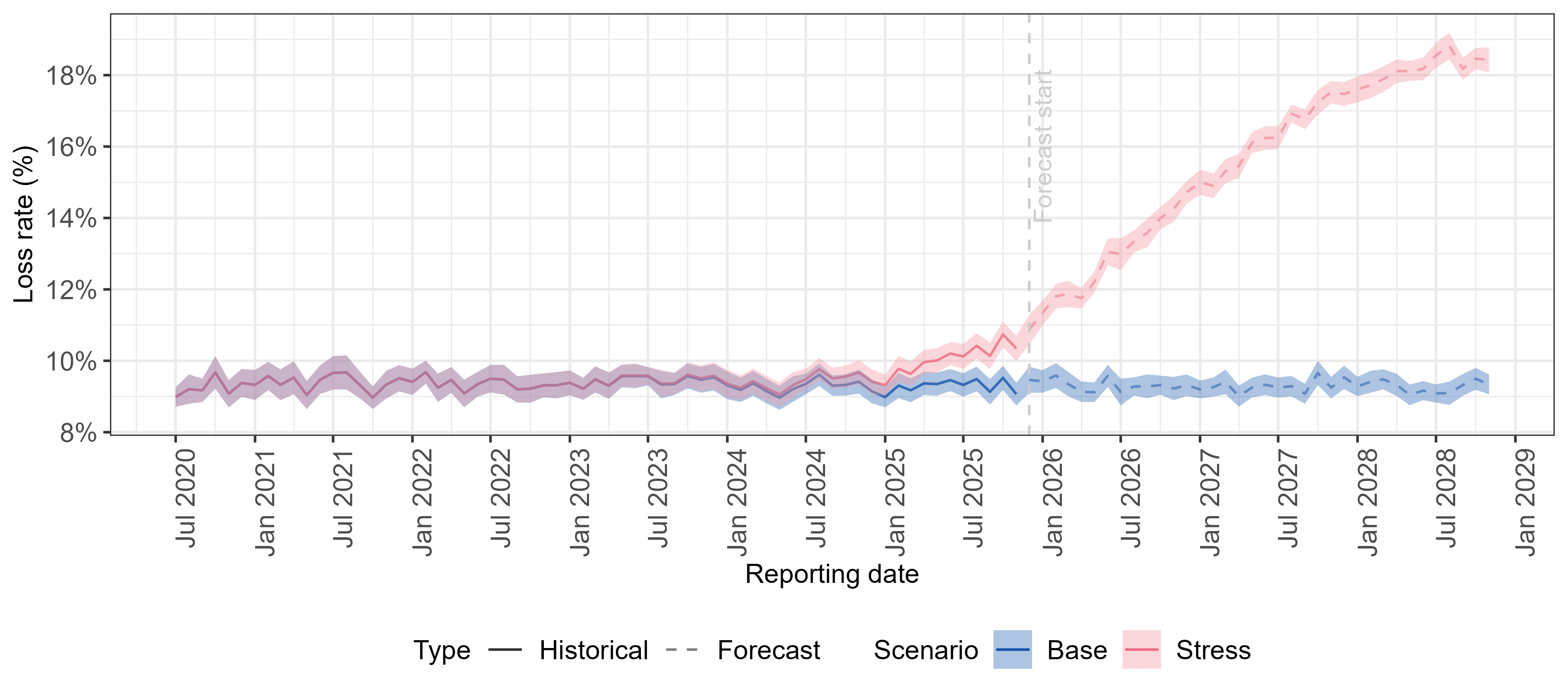}
\caption{A comparison between the average loss rates of the loan portfolios, across all simulation runs, for the baseline and stress scenario. The bands indicate the 95\% confidence interval.}
\label{fig:LossRate_Comparison}
\end{figure}

The increased number of defaulted loans in the stress scenario are subject to longer sojourn times, less likely repayments, and larger write-off amounts that ultimately culminate in an increased average loss rate across all simulations; as illustrated in \autoref{fig:LossRate_Comparison}. The rate increases substantially and gradually throughout the stress scenario, whilst maintaining volatility levels similar to the historic period. The approximately two-fold increase is again in line with expectations for a secured portfolio that is subject to an adverse event.

\section{Conclusion}
\label{sec:ch5}

Many stress-test frameworks exist within literature that banks can employ to perform stress tests as required by their respective central banks. However, none of these frameworks simultaneously address three critical aspects when stress testing a credit portfolio. Firstly, the correlation structure amongst the credit risk parameters, as used in estimating the credit risk, is commonly ignored such that the overall credit risk can potentially be misestimated. Secondly, a bank generally grants new loans during any period, regardless of the stress scenario, which affects the bank's total exposure and credit losses; a fact often ignored by most stress-testing approaches. Lastly, most stress-tests are conducted on a portfolio-level, which does not allow time-dependent and idiosyncratic loan-specific factors to be easily embedded.

To address these gaps, we present the EIDFAST-approach that focuses on generating risk-weighted and scenario-dependent receipts over time for each active and new loan in a given credit portfolio. By focussing on receipt-generation, our approach is \textit{integrated} since it embeds the well-known correlation structure amongst the credit risk parameters implicitly via the receipts. A loan production component is embedded within the broader forecasting framework, which ensures that the approach is also \textit{extendable}. Our approach is lastly \textit{dynamic} since it enables some of its parameters to be modelled as time-dependent and idiosyncratic loan-specific factors. To illustrate the application of the EIDFAST-approach, we simulated a portfolio over an historical period, subject to specific conditions. This portfolio was then forecast using the EIDFAST-approach over a three year forecast horizon under a baseline and a stress scenario. In fact, a Monte Carlo setup was used to simulate the loan portfolio 50 times for each of the baseline and stress scenarios, thereby enabling the degree of uncertainty in each scenario to be evaluated. The resulting portfolios are compared to one another using the 12-month default rate and loss rate, which are chosen as our two main risk metrics. The average default rate over the simulations of the baseline scenario was stable and hovered around a constant level. In contrast, the average default rate over the stress scenarios gradually increased 3.5-fold over the forecast horizon. Similarly, the average loss-rate over the baseline scenarios hovers around a constant level, though the volatility in the rate is higher than for the default rate. The volatility of the default rate was minuscule during each scenario, thereby indicating a high level of confidence in the forecasts. The average loss-rate rises gradually over the stress scenarios and is approximately twice that of the baseline scenarios by the end of the forecast horizon. The volatility of these average loss rates was higher than that of the average 12-month default rates, but was sufficiently small that the forecasts appear stable within the simulated setting. The increase in these risk metrics over the stress scenarios, relative to the baseline scenarios, affirm the ability of the EIDFAST-approach to produce credible and intuitive stress-test results, whilst addressing pertinent gaps in current stress-test frameworks.

Future research can explore the calibration of the EIDFAST-approach to real-world credit data, subject to a given baseline and stress scenario. This calibration will entail fitting various models to each of the sub-components within the loan-production and receipt-generation components. For the loan production component, each of the distributions used to simulate loan-attributes can be modelled as a function of various systematic effects, e.g., macroeconomic factors. Regarding the receipt generation component, loan-level models can be used to estimate the account-state transition probabilities of each loan such that idiosyncratic and systematic factors can be embedded as covariates. Furthermore, the payment probabilities $b$ and $b'$ need to be estimated through some loan- or portfolio-level model. Finally, the sojourn time and write-off rate distributions need to be calibrated, and can be done in a similar fashion to that of the loan production component's distributions.

Another avenue of future research may be to investigate and embed the correlation structure between the loan attributes within the loan production component. E.g., a bank may tend to disburse larger principal amounts to lower risk individuals, who will in turn be assigned lower annual interest rates. Such a relationship may even exist between the sojourn time and write-off rate of a written-off loan, and can therefore be embedded within the simulation of each of these receipt-generation components. In its current form, the EIDFAST-approach assumes that there are no partial repayments. This assumption may be unrealistic for certain portfolios, e.g., a retail credit card portfolio, but can be addressed by extending the EIDFAST-approach. Our approach can also be expanded by incorporating loan re-advances or re-capitalisations, which may be highly prevalent within revolving products such as credit cards and revolving loans. Lastly, the receipts generated for a loan may be weighted by its associated transition probabilities, which better reflects the uncertainty of these receipt forecasts; see for instance \citet{djeundje2025}. In conclusion, our proposed EIDFAST-approach simultaneously addresses many pertinent gaps in existing stress-testing frameworks and can be expanded and enhanced such that it can be applied to a wide variety of loan portfolios.

\appendix
\section{Appendix: Parametrising the EIDFAST-approach for a baseline scenario}
\label{app:app01}

\begin{figure}[H]
\begin{subfigure}{.5\textwidth}
  \centering
    \caption{New loans}
  \includegraphics[width=1\linewidth]{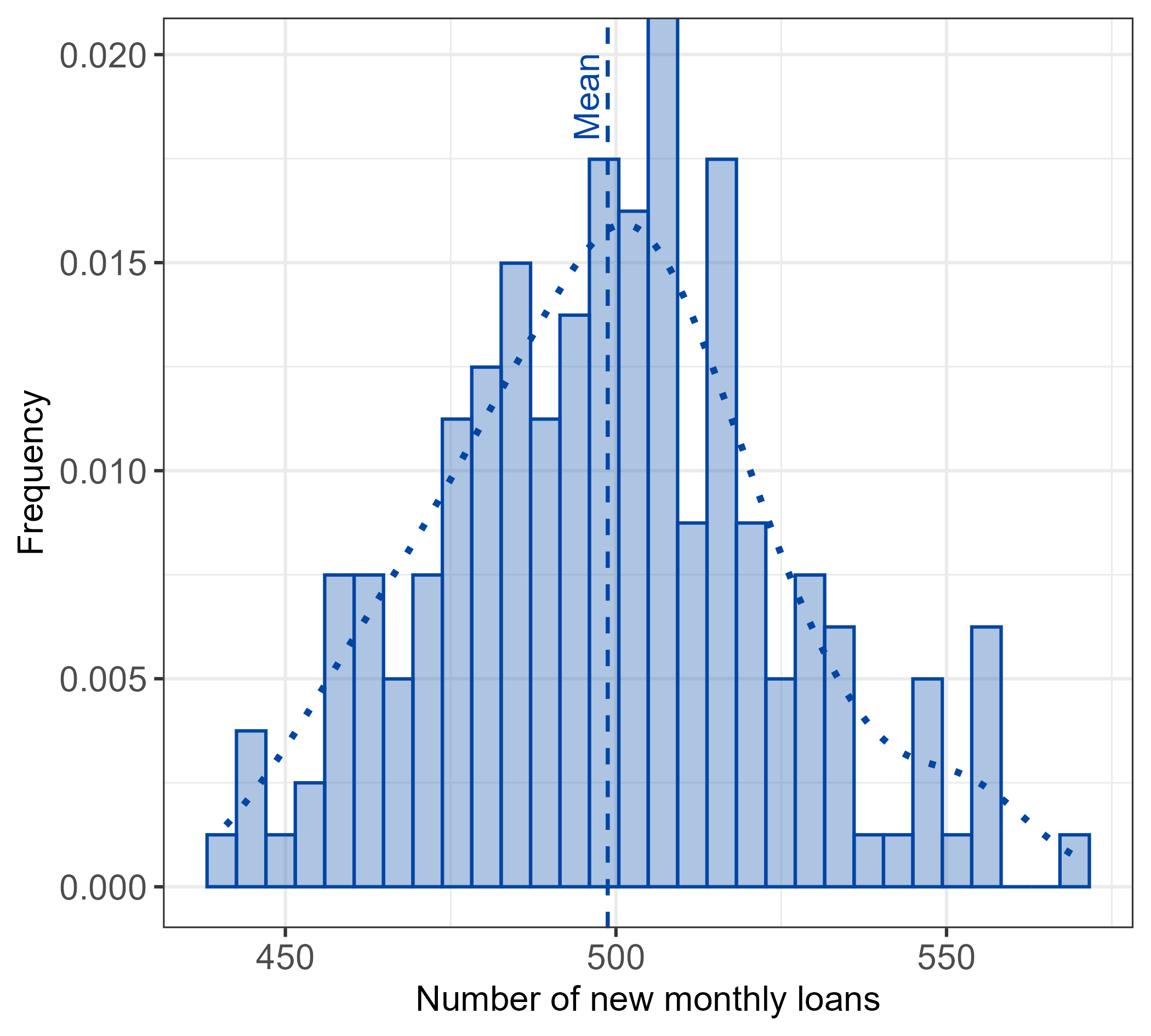}
  \label{fig:New_loans_dist}
\end{subfigure}
\begin{subfigure}{.5\textwidth}
  \centering
  \caption{Principal amounts}
  \includegraphics[width=1\linewidth]{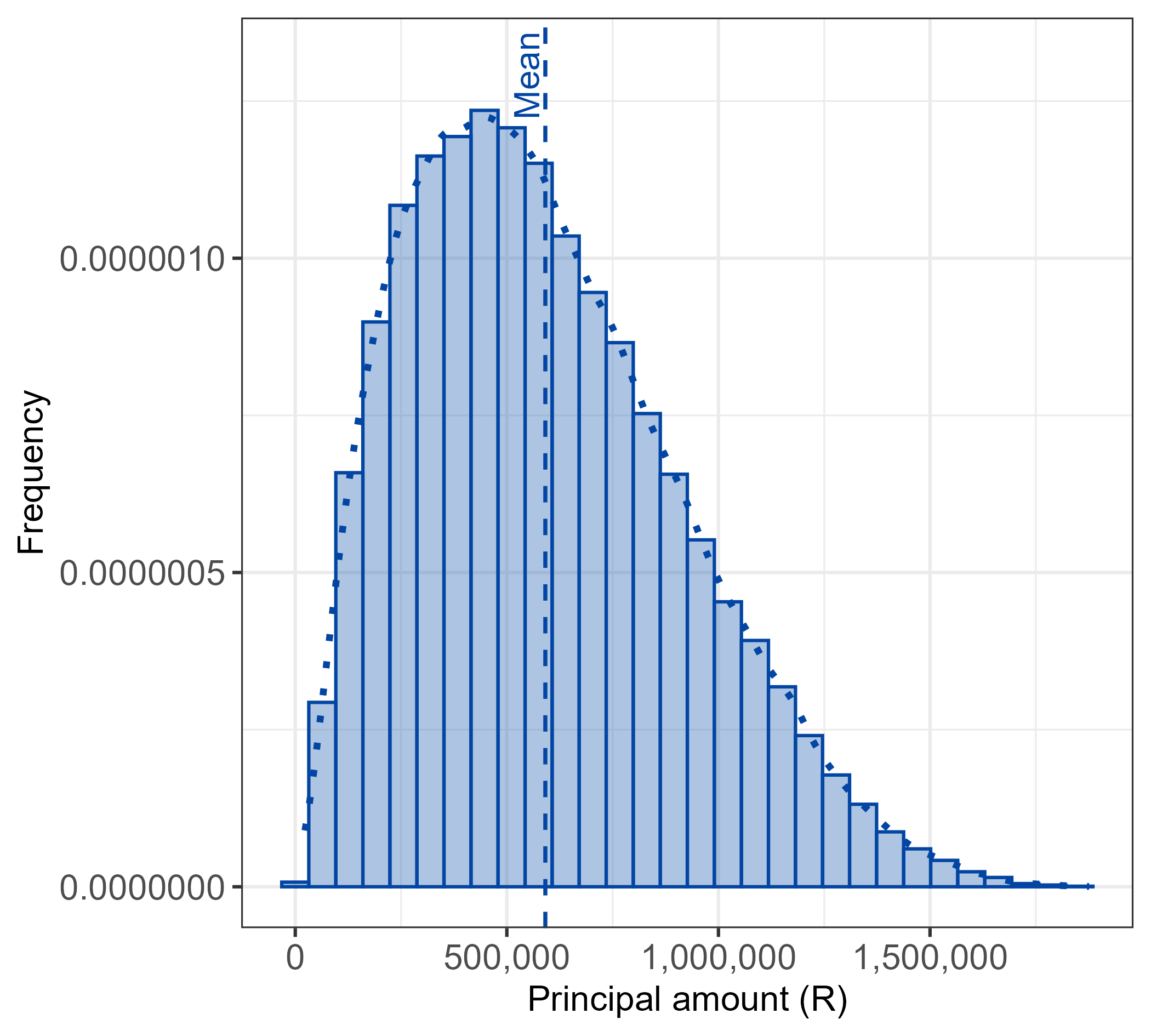}
  \label{fig:Principal_dist}
\end{subfigure}
\begin{subfigure}{.5\textwidth}
  \centering
    \caption{Terms}
  \includegraphics[width=1\linewidth]{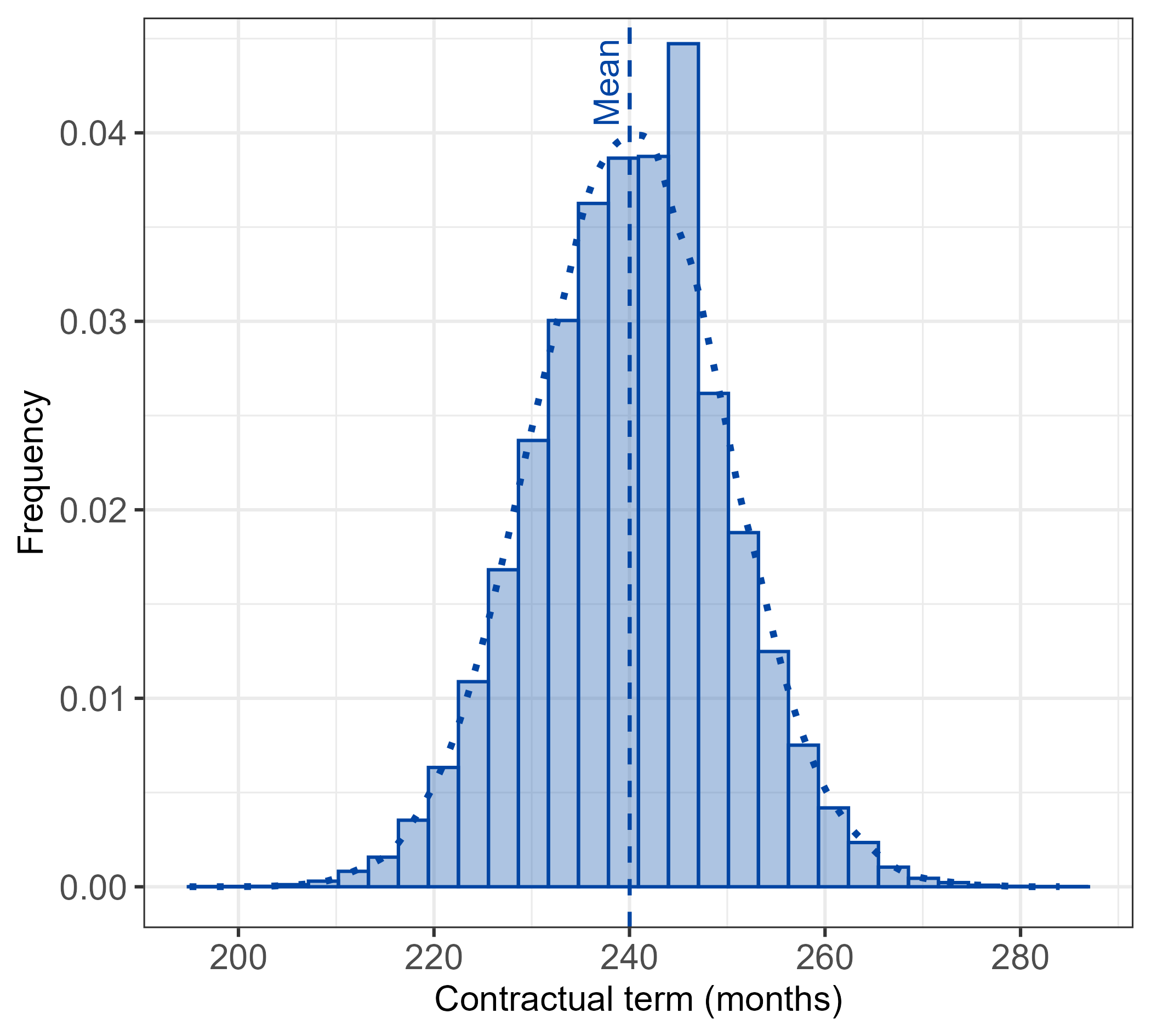}
  \label{fig:Term_dist}
\end{subfigure}
\begin{subfigure}{.5\textwidth}
  \centering
    \caption{Nominal interest rates}
  \includegraphics[width=1\linewidth]{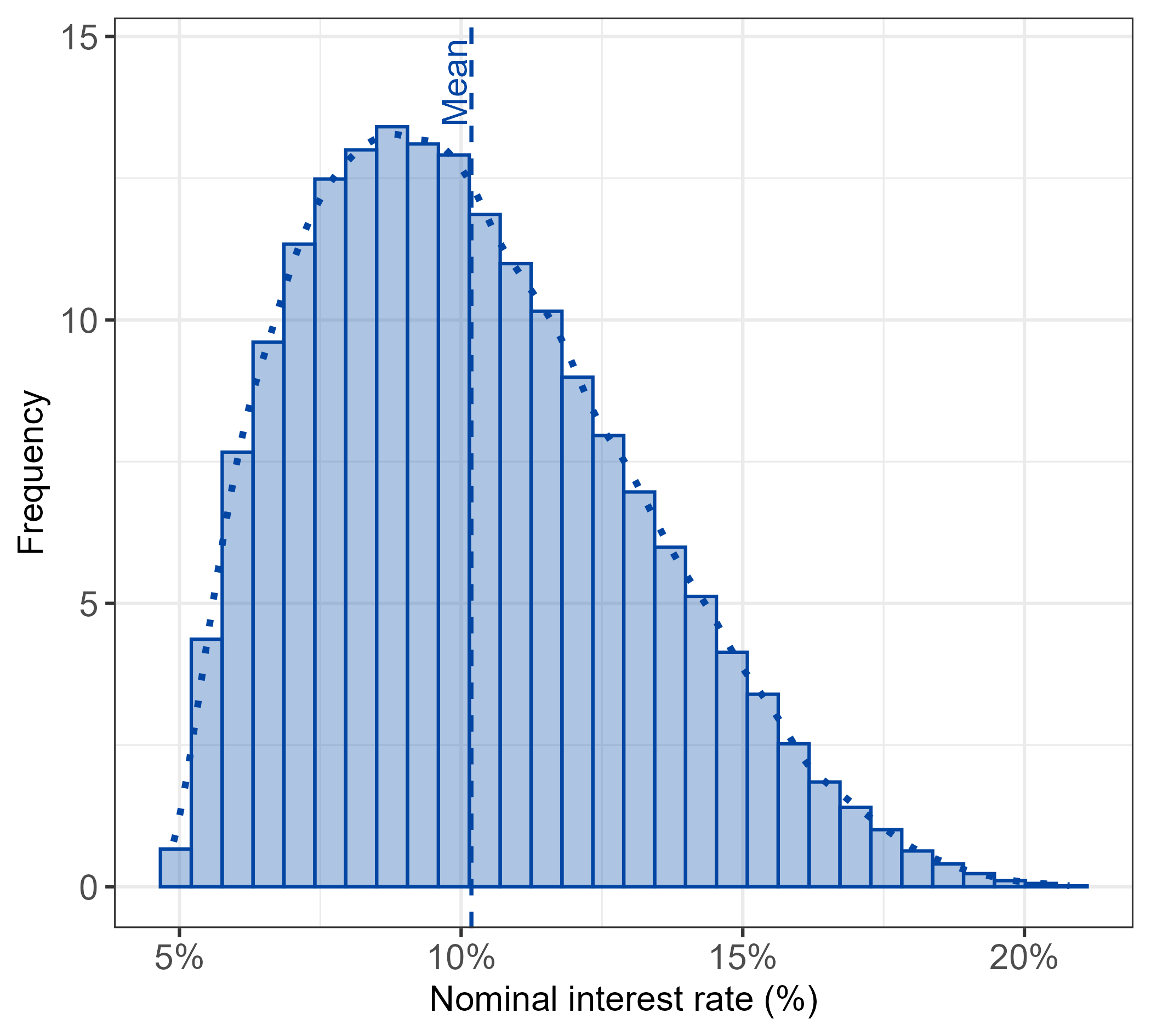}
  \label{fig:Int_dist}
\end{subfigure}
\caption{Histogram and empirical densities of loan characteristics within the loan production component.}
\label{fig:dist_historical}
\end{figure}

\begin{figure}[H]
\begin{subfigure}{.5\textwidth}
  \centering
  \caption{Sojourn times}
  \includegraphics[width=1\linewidth]{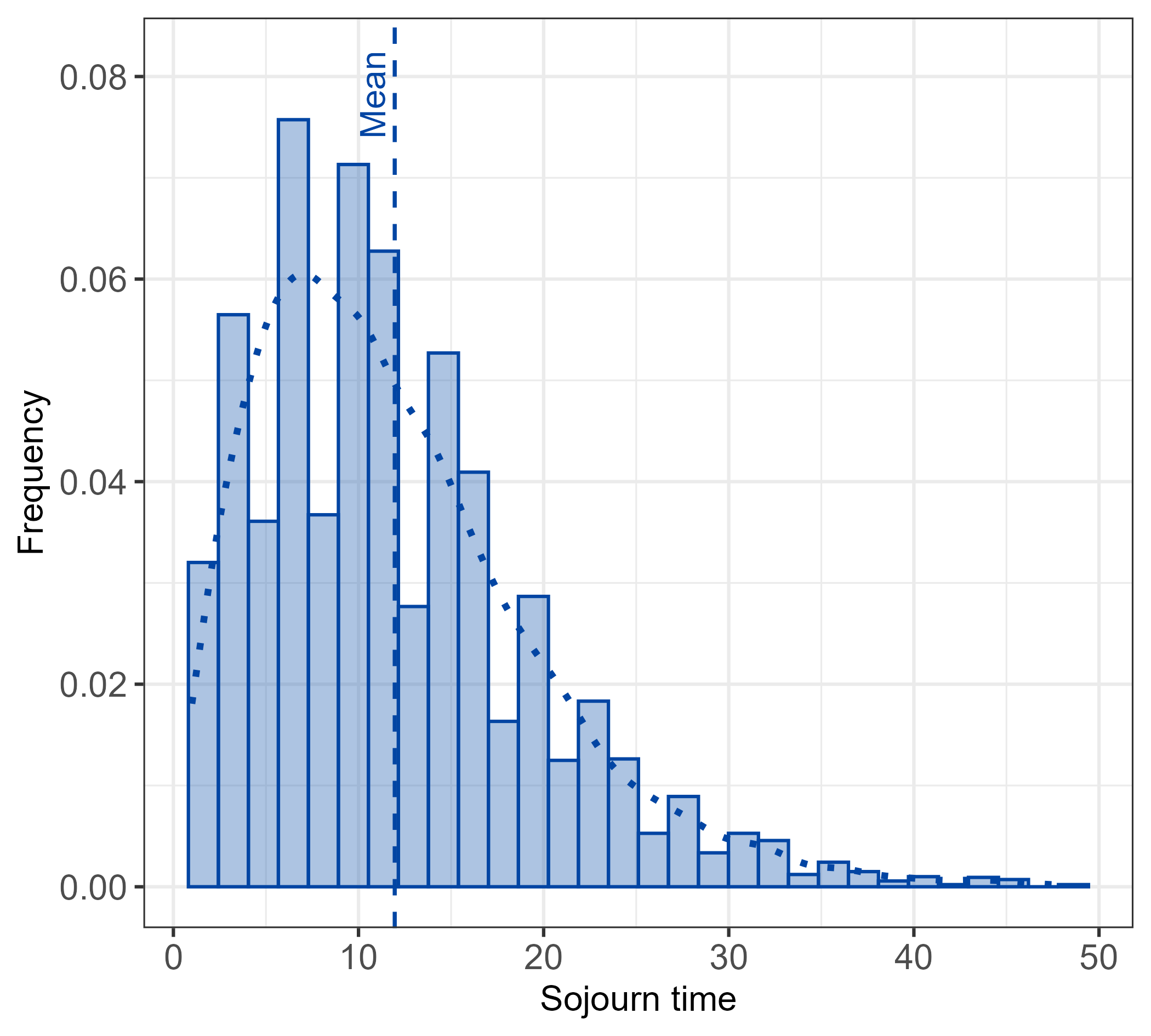}
  \label{fig:Sojourn_times_dist}
\end{subfigure}%
\begin{subfigure}{.5\textwidth}
  \centering
  \caption{Write-off rates}
  \includegraphics[width=1\linewidth]{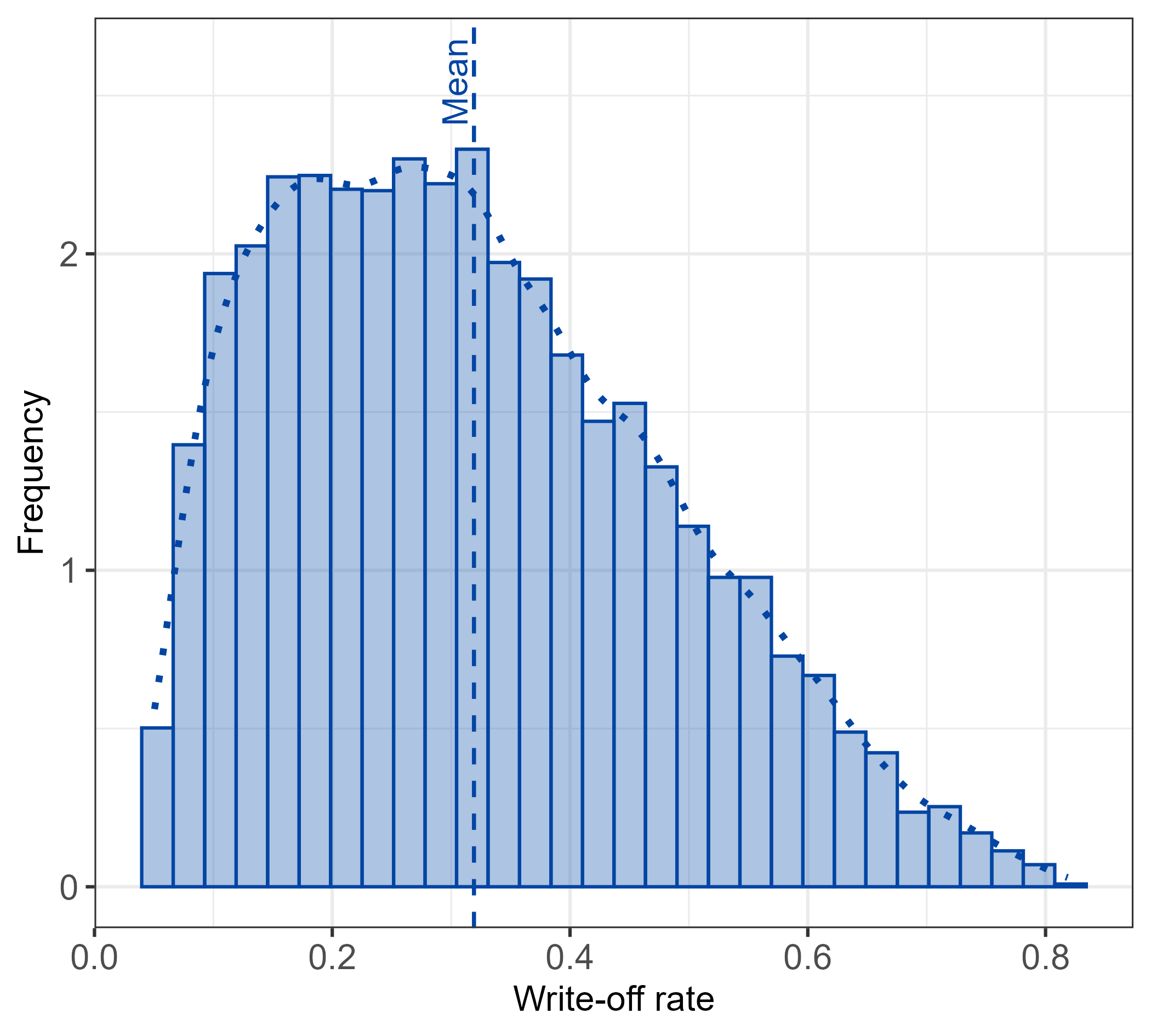}
  \label{fig:Woff_rates_dist}
\end{subfigure}
\caption{Histogram and empirical densities of sojourn times and write-off rates.}
\label{fig:dist2}
\end{figure}

The distributions of the sub-components used in the EIDFAST-approach with respect to the baseline scenario are illustrated in this appendix. Firstly, we show the distributions in \autoref{fig:dist_historical} of the number of new loans extended at a single period, together with that of each new loan's associated principal amount, contractual term, and nominal (annual) interest rate. Secondly, the distributions of the sojourn times and write-off rates assigned to the loans destined for write-off are illustrated in \autoref{fig:dist2}.

\section{Appendix: The evolution of framework components subject to specific scenarios}
\label{app:app02}
As detailed in \autoref{sec:ch4}, a loan portfolio is simulated using the EIDFAST-approach 50 times for both the baseline and stress scenario. The effects of these scenarios on the evolution of the loan production and receipt forecast components, with respect to the stress scenario, are respectively summarised in \autoref{tbl:parameters_stress_summary_loan_prod} and \autoref{tbl:parameters_stress_summary_receipts} over the forecast horizon. In this appendix, a detailed analysis is provided of each sub-component by comparing its distribution between scenarios. The evolution of the sub-component's distribution over all simulations, and with respect to each scenario, is then further analysed.
As explained in \autoref{sec:ch3.2}, these evolutions of each sub-component are assumed to be linear over the forecast horizon and are enforced by applying \autoref{eq:parameter_stress}.
Moreover, z-scores are used to construct 95\% confidence intervals for each sub-component across all simulations runs and for each scenario to assess the uncertainty.

\begin{figure}[H]
\centering
\begin{subfigure}{\textwidth}
  \centering
    \caption{Distributions of new monthly loans}\includegraphics[width=\textwidth,height=8.25cm]{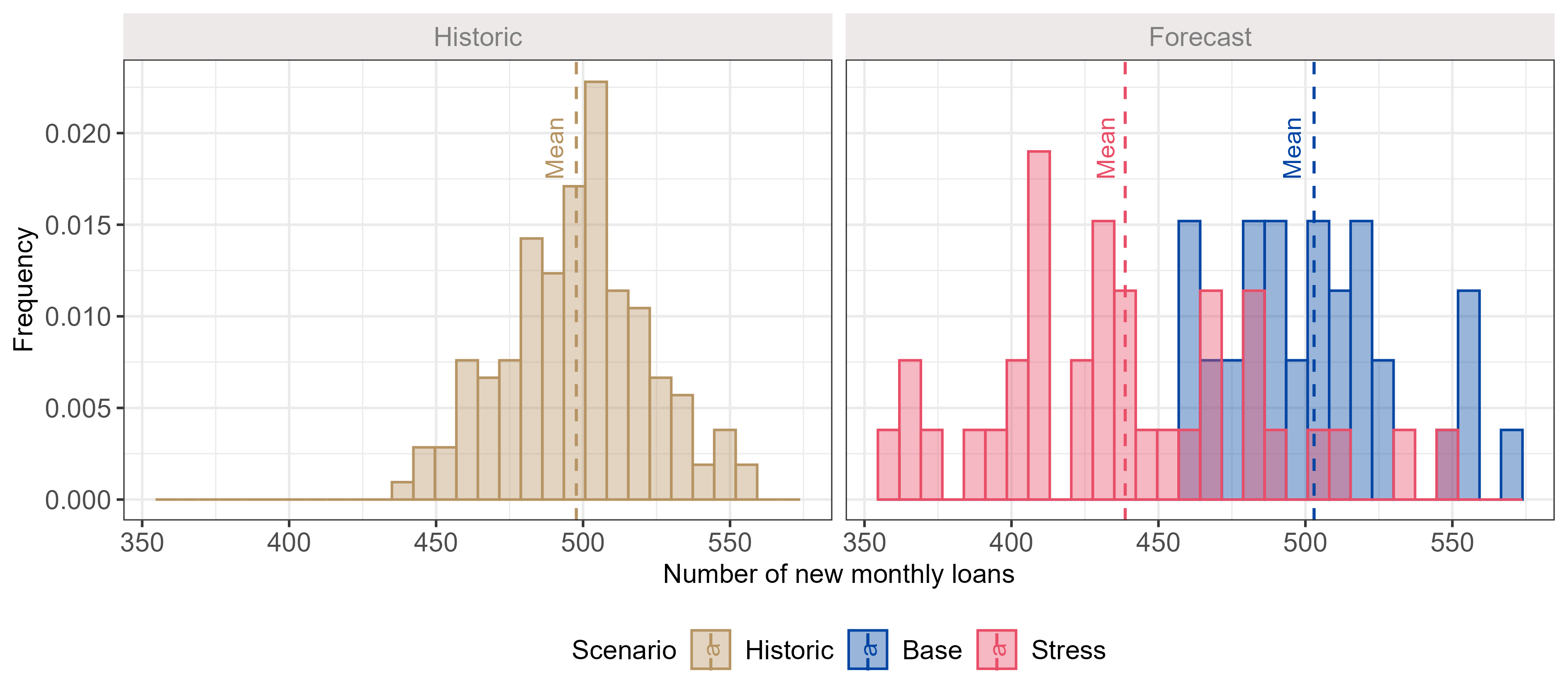}

  \label{fig:NewLoans_Distribution_Comp}
\end{subfigure}
\begin{subfigure}{\textwidth}
  \centering
  \caption{Average number of new monthly loans over time}  \includegraphics[width=\textwidth,height=8.25cm]{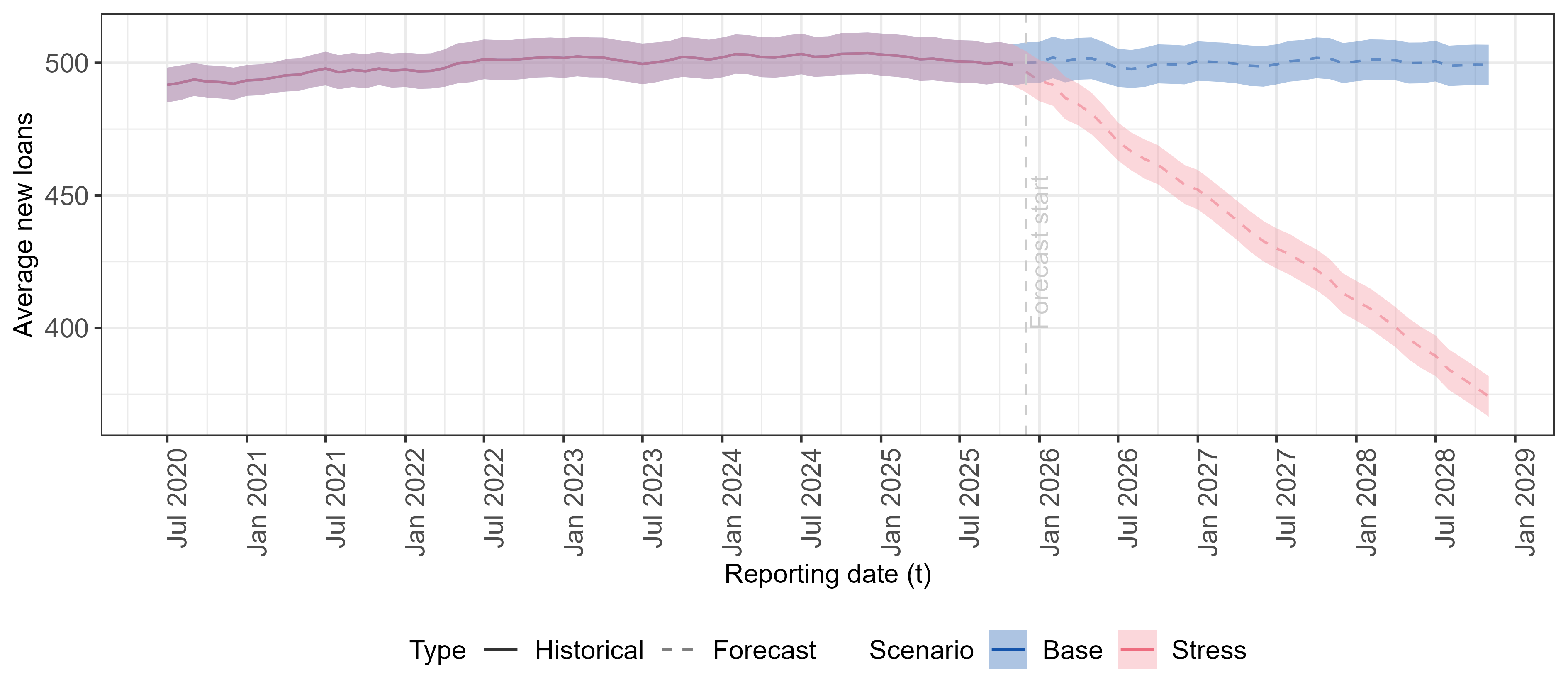}

  \label{fig:NewLoans_Ave_Comp}
\end{subfigure}
\caption{A comparison between the number of new loans disbursed across the historical and forecast periods, with respect to the baseline and stress scenarios. In panel \textbf{(a)}, the distributions of new monthly loan volumes are shown per scenario and period type. The average number of new monthly loans is shown in panel \textbf{(b)} over time, with 95\% confidence intervals.}
\label{fig:NewLoans_Comparison}
\end{figure}

\begin{figure}[H]
\centering
\begin{subfigure}{\textwidth}
  \centering
    \caption{Distributions of principal amounts}
    \includegraphics[width=\textwidth,height=8.25cm]{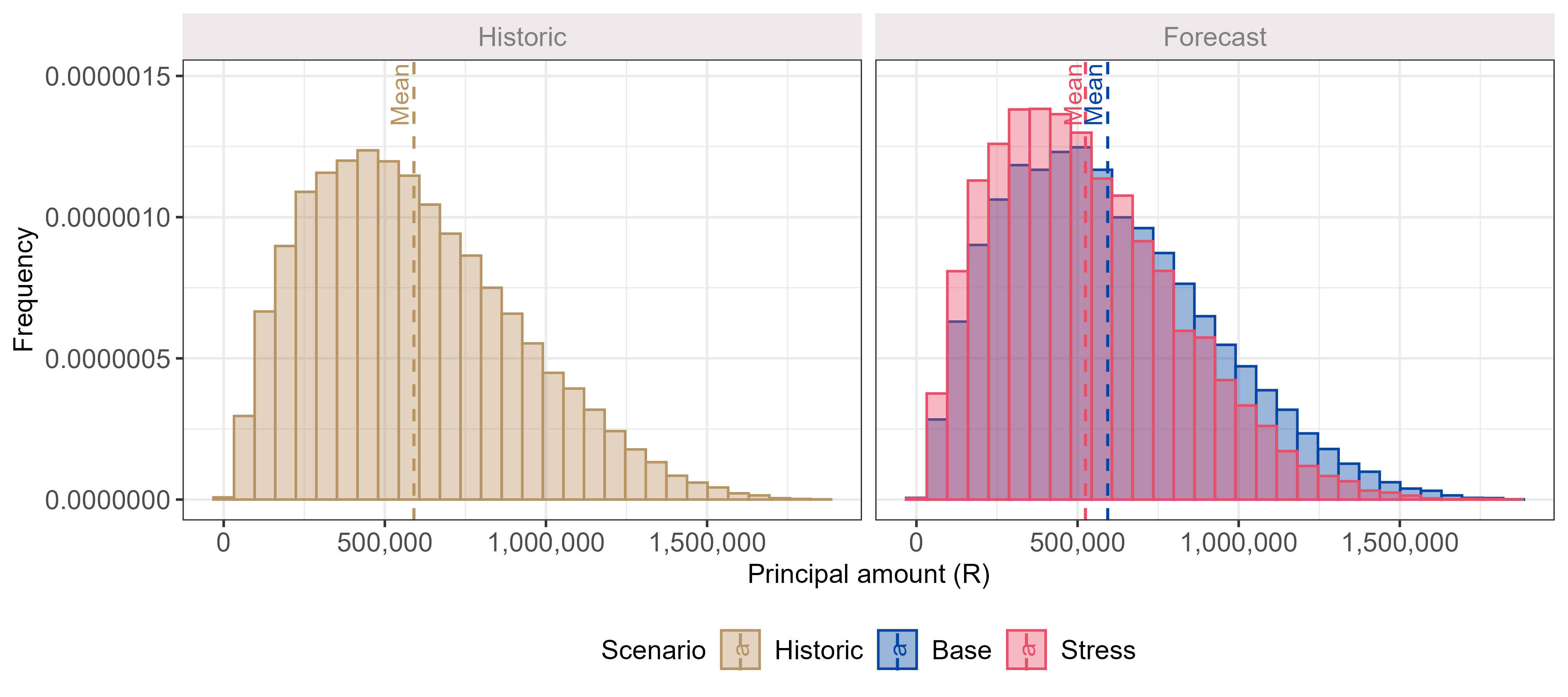}
  \label{fig:Principals_Distribution_Comp}
\end{subfigure}
\begin{subfigure}{\textwidth}
  \centering
    \caption{Average principal amount over time}
    \includegraphics[width=\textwidth,height=8.25cm]{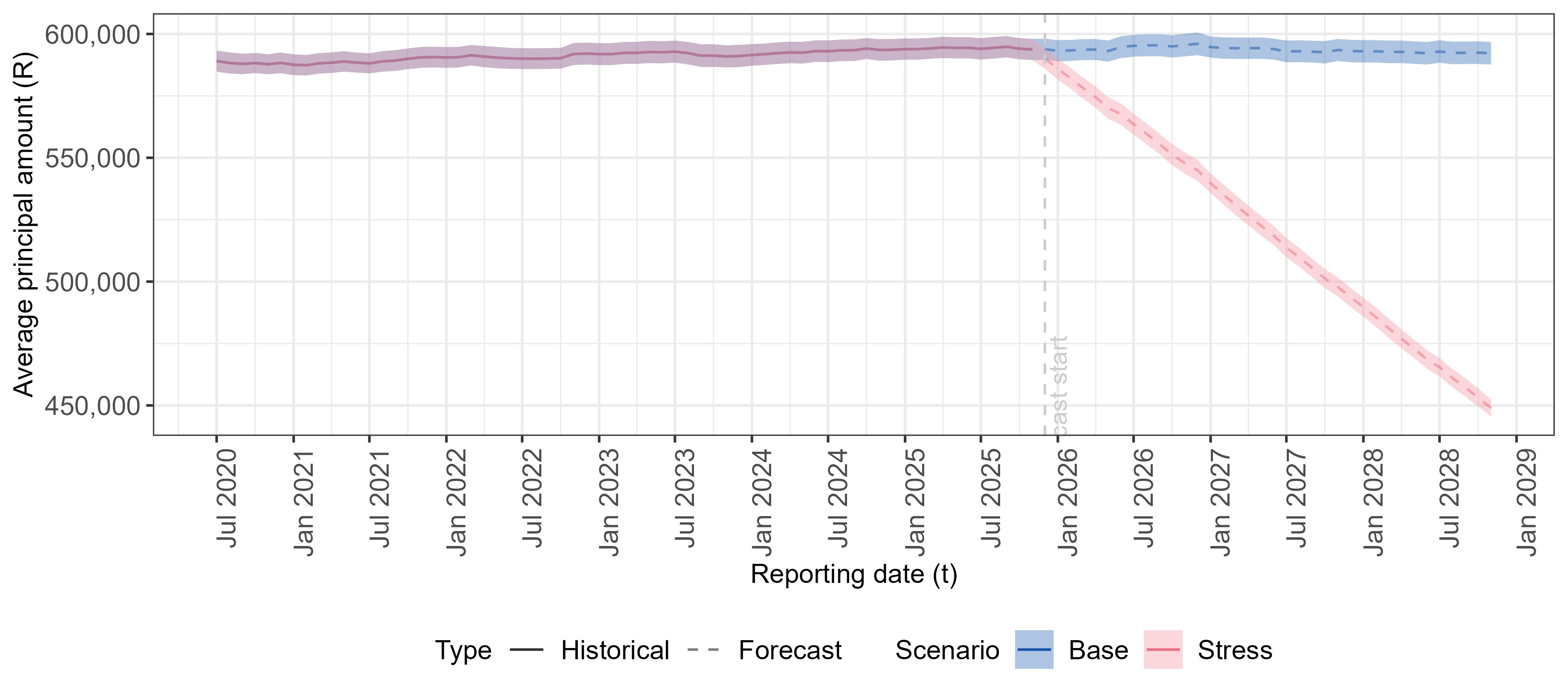}
  \label{fig:Principals_Ave_Comp}
\end{subfigure}
\caption{A comparison between the principal amounts disbursed across the historical and forecast periods, with respect to the baseline and stress scenarios. The graph design follows that of \autoref{fig:NewLoans_Comparison}.}
\label{fig:Principals_Comparison}
\end{figure}

\begin{figure}[H]
\centering
\begin{subfigure}{\textwidth}
  \centering
    \caption{Distributions of annual interest rates assigned}
    \includegraphics[width=\textwidth,height=8.25cm]{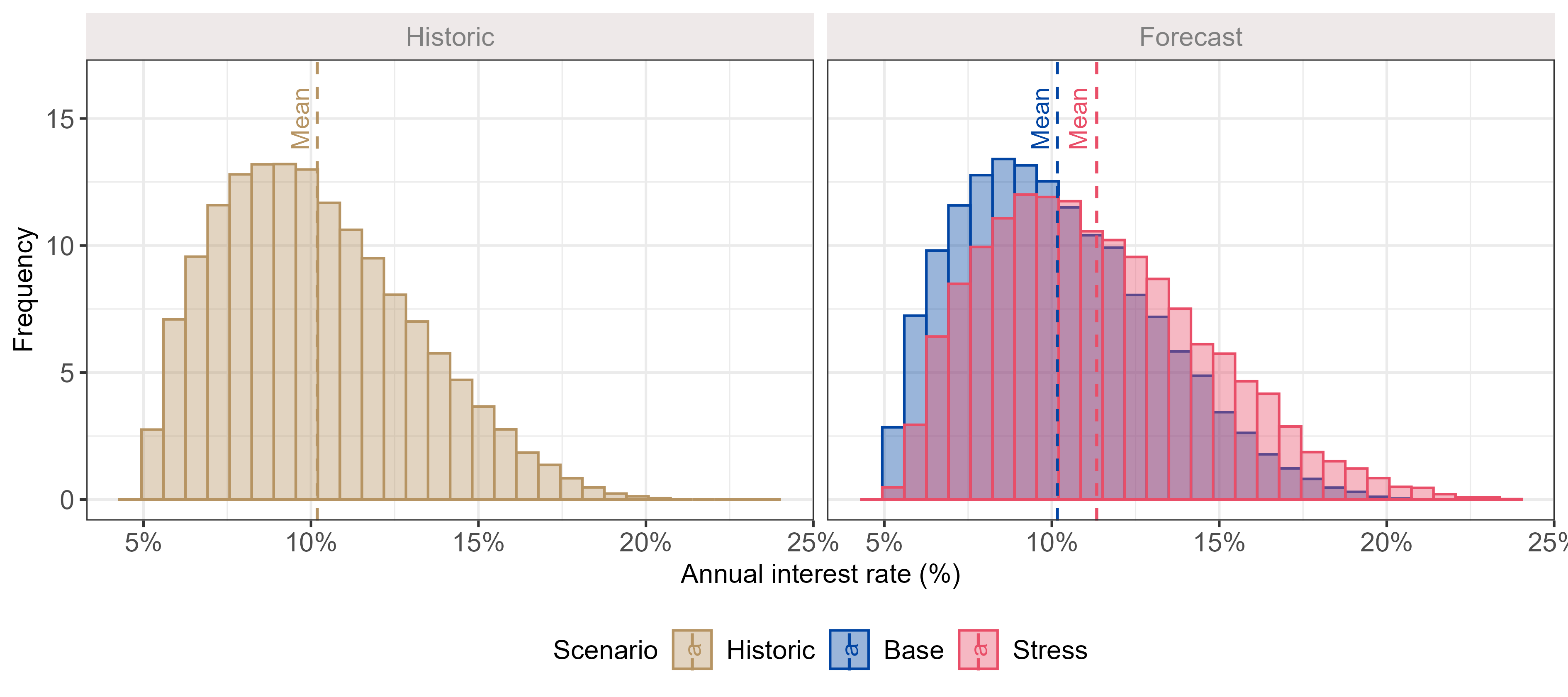}
  \label{fig:AnnualInterest_Rate_Distribution_Comp}
\end{subfigure}
\begin{subfigure}{\textwidth}
  \centering
    \caption{Average annual interest rate over time}
    \includegraphics[width=\textwidth,height=8.25cm]{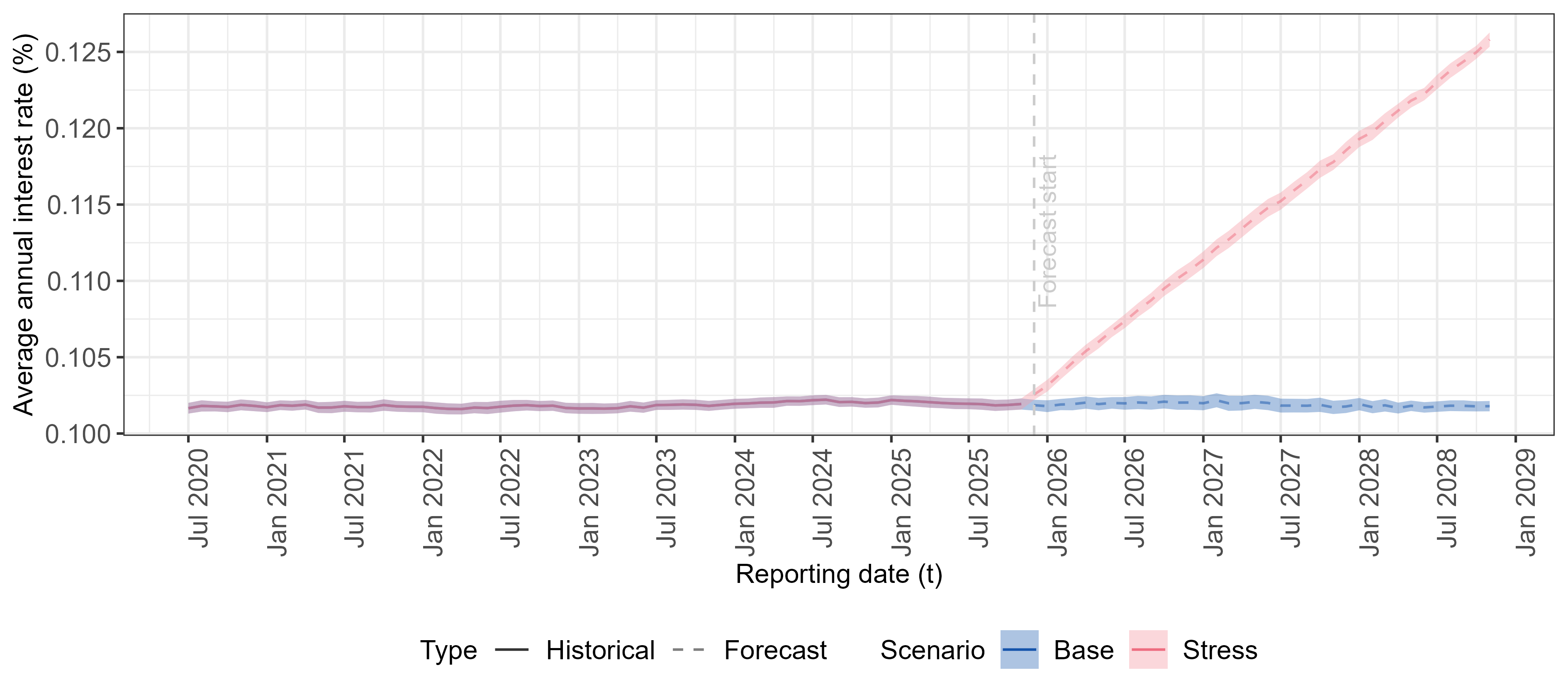}
  \label{fig:AnnualInterest_Rate_Ave_Comp}
\end{subfigure}
\caption{A comparison between the annual interest rates assigned to new loans across the historical and forecast periods, with respect to the baseline and stress scenarios. The graph design follows that of \autoref{fig:NewLoans_Comparison}.}
\label{fig:Annual_InterestRate_Comparison}
\end{figure}

The sub-components within the loan production component are analysed and compared between each of the two scenarios as follows.
Consider the number of new monthly loans disbursed over the historical and forecast period for the baseline and stress scenario for a single simulation, as illustrated in \autoref{fig:NewLoans_Distribution_Comp}. In the stress scenario, the resulting distribution shifts to the left such that the average number of new monthly loans decreases overall. This contraction is reflected in \autoref{fig:NewLoans_Ave_Comp}, which also shows that the average number of new monthly loans decreases steadily over the forecast horizon.
The associated principal amount assigned to each new loan is also reduced in the stress scenario, as seen in the shift in its distribution to the left in \autoref{fig:Principals_Distribution_Comp}. The average principal amount also decreases over the forecast horizon for all simulations; see \autoref{fig:Principals_Ave_Comp}.
Moreover, the distribution of the annual interest rates assigned to new loans shifts to the right and is shown in \autoref{fig:AnnualInterest_Rate_Distribution_Comp}. The average annual interest rate assigned to new loans within each simulation increases steadily over the forecast horizon, for the stress scenario, as seen in \autoref{fig:AnnualInterest_Rate_Ave_Comp}.
We believe that all of these phenomena attest to stricter lending criteria that are enforced during a stress period.

\begin{figure}[H]
\centering
\begin{subfigure}{\textwidth}
  \centering
  \caption{Distributions of sojourn times}
  \includegraphics[width=\textwidth,height=8.25cm]{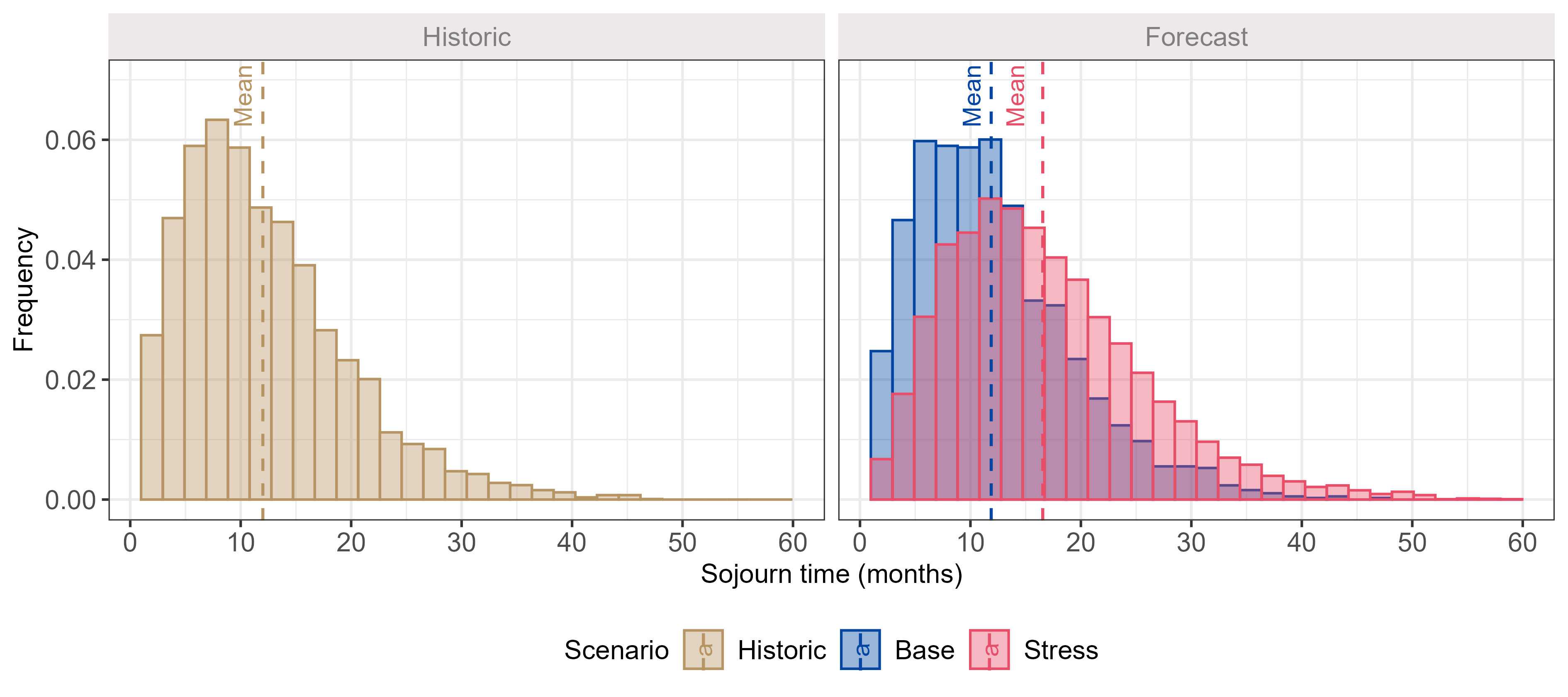}
  \label{fig:SojournTime_Distribution_Comp}
\end{subfigure}
\begin{subfigure}{\textwidth}
  \centering
  \caption{Average sojourn times over time}
  \includegraphics[width=\textwidth,height=8.25cm]{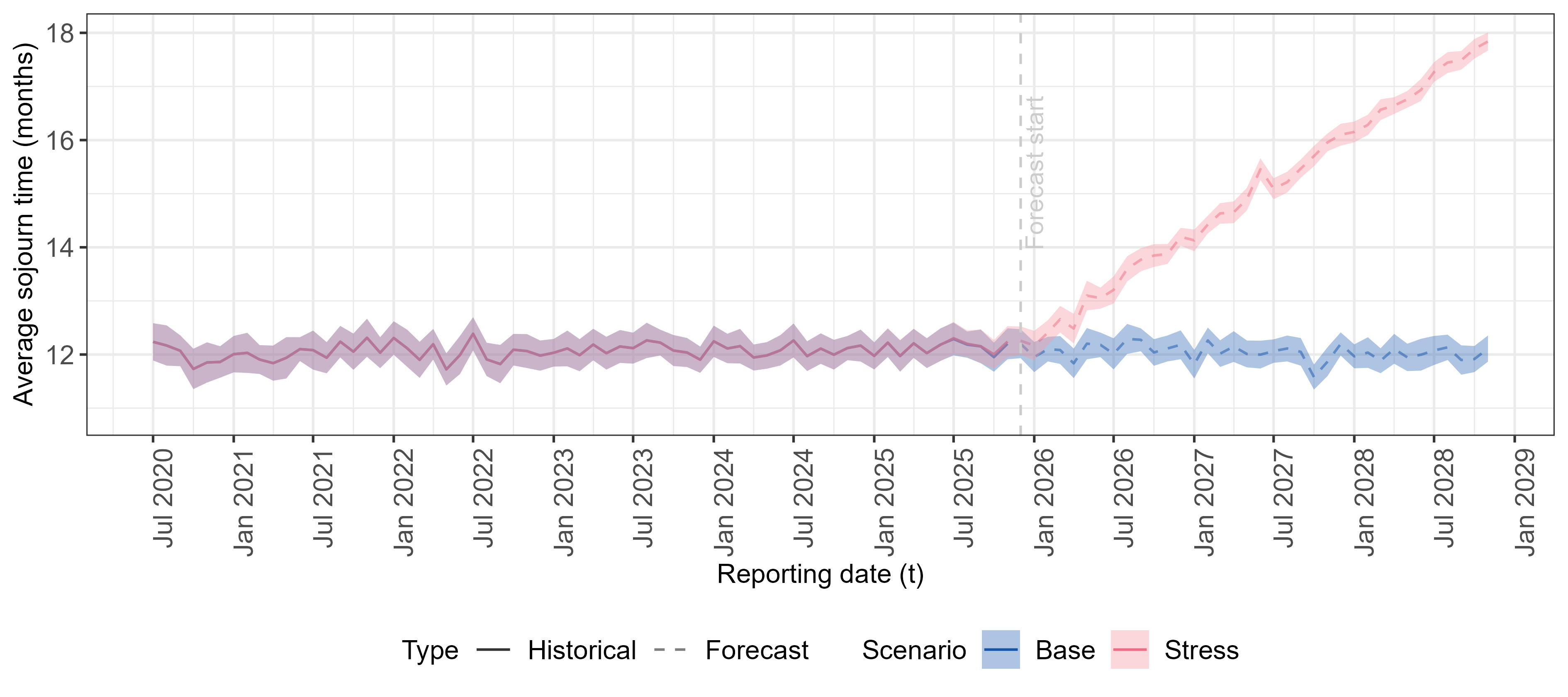}
  \label{fig:SojournTime_Ave_Comp}
\end{subfigure}
\caption{A comparison between the sojourn times assigned to loans destined to write-off across the historical and forecast periods, with respect to the baseline and stress scenarios.  The graph design follows that of \autoref{fig:NewLoans_Comparison}.}
\label{fig:SojournTime_Comparison}
\end{figure}

\begin{figure}[H]
\centering
\begin{subfigure}{\textwidth}
  \centering
  \caption{Distributions of write-off rates}
  \includegraphics[width=\textwidth,height=8cm]{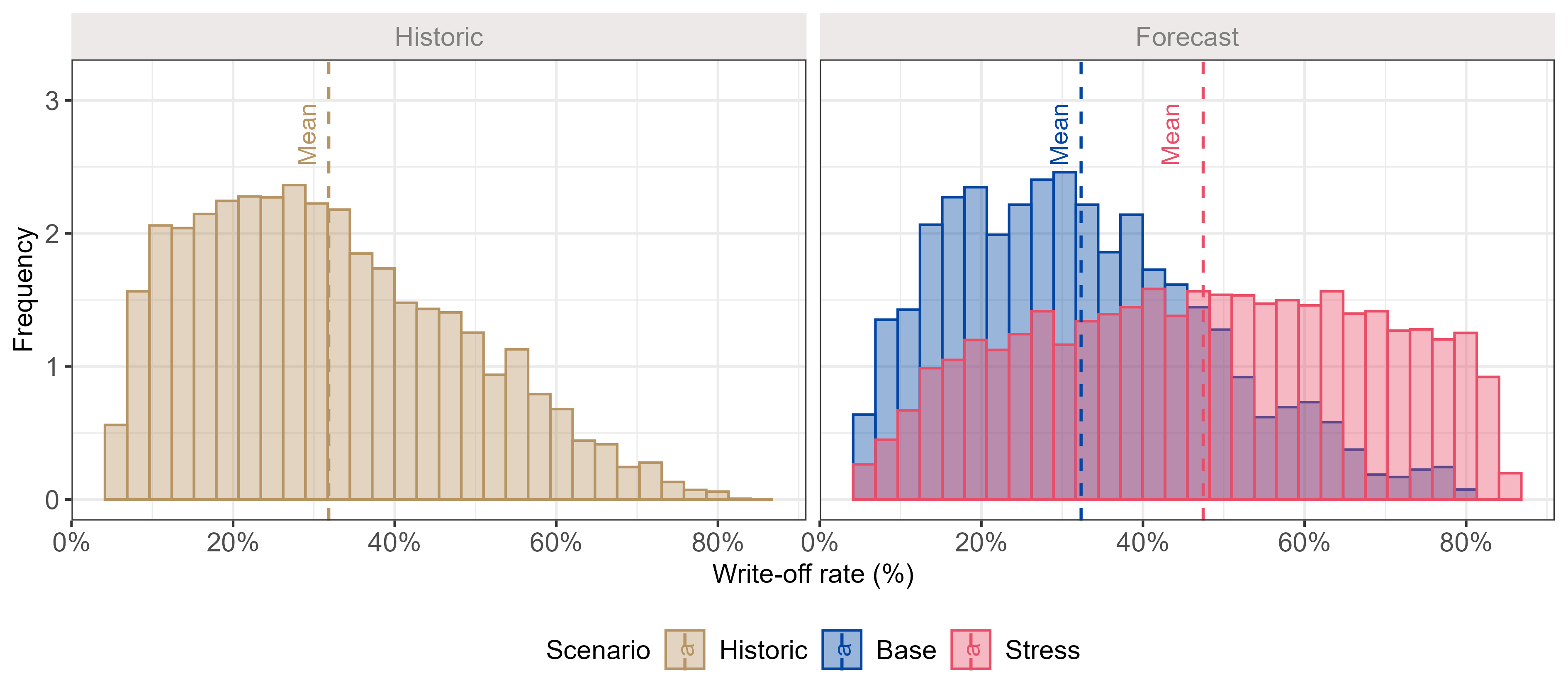}
  \label{fig:WOffRate_Distribution_Comp}
\end{subfigure}
\begin{subfigure}{\textwidth}
  \centering
  \caption{Average write-off rate over time}\includegraphics[width=\textwidth,height=8.25cm]{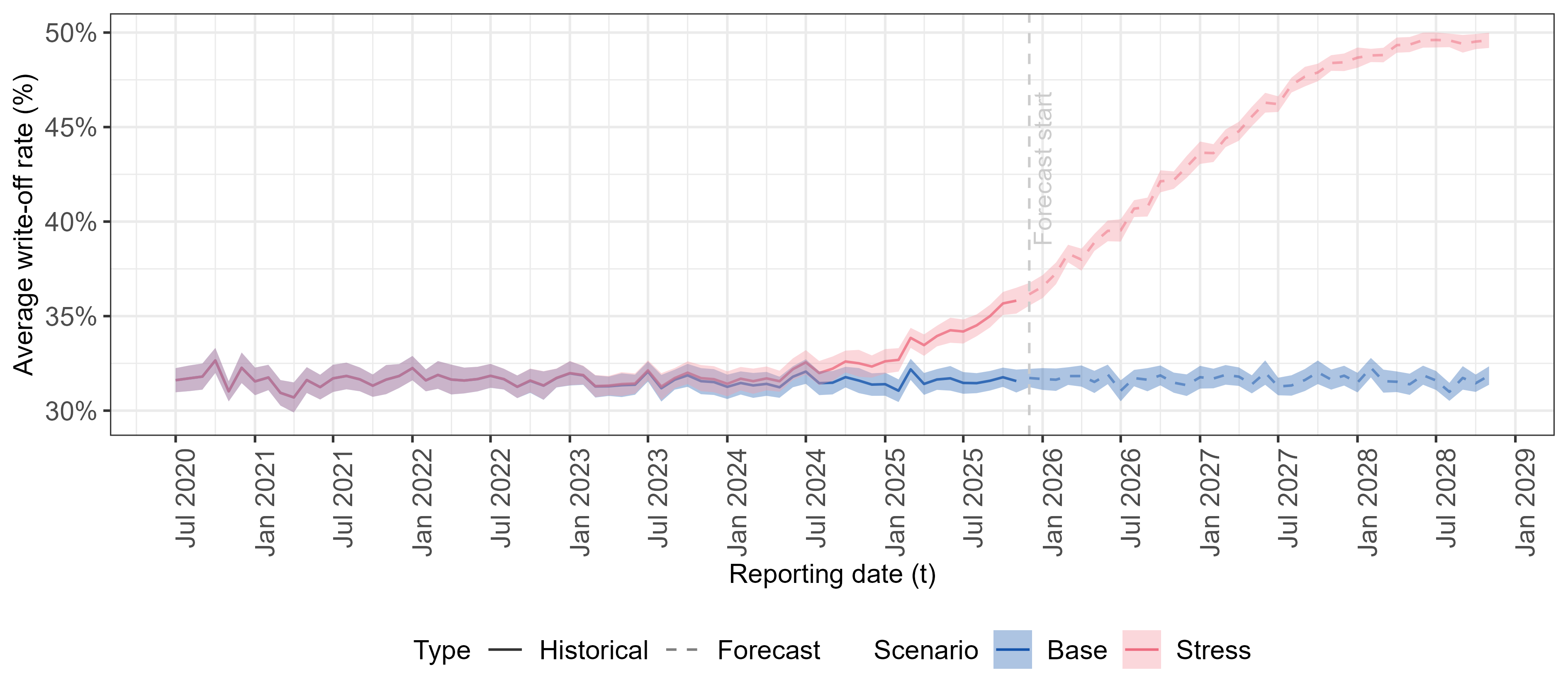}
  \label{fig:WOff_Rate_Ave_Comp}
\end{subfigure}
\caption{A comparison between the write-off rates assigned to written-off loans across the historical and forecast periods, with respect to the baseline and stress scenarios. The graph design follows that of \autoref{fig:NewLoans_Comparison}}
\label{fig:WOff_Rate_Comparison}
\end{figure}

We now focus on the receipt forecasting sub-components, n.l., the sojourn times and write-off rates assigned to loans destined for write-off.
The distribution of sojourn times shifts to the right in the stress scenario, as seen in \autoref{fig:SojournTime_Distribution_Comp}. The associated average sojourn time, illustrated in \autoref{fig:SojournTime_Ave_Comp}, increases over the forecast horizon in the stress scenario.
Similarly, \autoref{fig:WOff_Rate_Comparison} shows the distribution of the write-off rates shifting to the right in the stress scenario, whilst the average write-off rate increases over the forecast horizon.

\section{Appendix: Portfolio risk metrics}
\label{app:app03}

The 12-month default rate and the loss rate are chosen as the portfolio-level risk metrics to assess the credit risk of our simulated loan portfolios. Estimating these metrics requires account-level observations of each active loan $i=1,\dots,N$ within the loan portfolio of size $N$ over the portfolio's lifetime. 
To calculate the portfolio-level default rate, first define `default' as a binary event $z_{i,t} \in \{0,1\}$, 
where one indicates default and zero indicates non-default at time $t$.
Let $\mathcal{D}_{t}$ 
be the set of all loans $i$ at risk of default at $t$. 
The 12-month default rate at time $t$ 
is then defined as
\begin{equation} \label{eq:default_rate_conditional}
    \Gamma(t) =  \frac{1}{|\mathcal{D}_{t} |}\sum_{i \, \in \, \mathcal{D}_{t}} \max(z_{i,t+1},\dots,z_{i,t+12})  \, .
\end{equation}


In estimating the portfolio-level LGD, consider a single loan $j$ that defaulted at time $t$ and that is written-off at time $t_{w}$. During this loan's workout process $s \in \{t+1,...,t_{w}\}$, collection efforts are likely made that yield receipts $R_{j,s}$. 
The realised loss $l_j(t)$ of $j$ at $t$ 
is expressed as the proportion of the loan's balance at default, $B_{j,t}$
, that is ultimately written-off after accounting for its receipts, i.e.,
\begin{equation}
    l_{j}(t) = 1 - \frac{\sum_{s=t+1}^{t_{j,w}}R_{j,s} \cdot v_{j,s}}{B_{j,t}} \, ,
\end{equation}
where $v_{j,s}$ is a factor that discounts $R_{j,s}$ to $t$ using the loan's interest rate. Next, let $\mathcal{L}_{t}$ be the set of all loans that defaulted at $t$.
The portfolio-level realised loss rate $l(t)$ at $t$ is estimated by averaging the realised loss for each defaulted loan at $t$, i.e., \begin{equation} \label{eq:Realised_Loss_Rate2}
    l(t) = \frac{1}{|\mathcal{L}_{t}|} \sum_{j \in \mathcal{L}_{t}} l_j(t) \, .
\end{equation}




\singlespacing
\printbibliography 
\onehalfspacing



\end{document}